\documentclass[twoside,english,prd,aps,amsmath,amssymb,reprint,pra,superscriptaddress,longbibliography]{revtex4-2}
\usepackage[T1]{fontenc}
\usepackage[latin9]{inputenc}
\usepackage{color}
\usepackage{textcomp}
\usepackage{amsmath}
\usepackage{amssymb}
\usepackage{graphicx}
\usepackage{subscript}

\makeatletter

\providecommand{\tabularnewline}{\\}

\usepackage{babel}
\usepackage{babel}

\makeatother

\usepackage{babel}
\begin{document}
\title{Quality of Service in \textcolor{black}{aggregated} quantum networks}
\author{Nicolo Lo Piparo}
\email{nicopale@gmail.com}

\affiliation{Okinawa Institute of Science and Technology Graduate University, 1919-1
Tancha, Onna-son, Okinawa, 904-0495, Japan.}
\author{William J. Munro}
\affiliation{Okinawa Institute of Science and Technology Graduate University, 1919-1
Tancha, Onna-son, Okinawa, 904-0495, Japan.}
\affiliation{National Institute of Informatics, 2-1-2 Hitotsubashi, Chiyoda-ku,
Tokyo 101-8430, Japan.}
\author{Kae Nemoto}
\affiliation{Okinawa Institute of Science and Technology Graduate University, 1919-1
Tancha, Onna-son, Okinawa, 904-0495, Japan.}
\affiliation{National Institute of Informatics, 2-1-2 Hitotsubashi, Chiyoda-ku,
Tokyo 101-8430, Japan.}
\begin{abstract}
\textcolor{black}{Future quantum networks will enable the interconnection
of multiple users distributed across vast geographic distances. Due
to these large separations and limited physical resources, communication
will often rely on multi-path routing strategies, where physical resources
are distributed across channels of varying lengths and delivered to
end users. Efficient long-distance quantum communication therefore
requires optimizing the allocation of these resources across available
paths. In this work, we introduce a performance-oriented approach
to quantum network routing by extending the classical concept of Quality
of Service (QoS) to the context of multi-path quantum resources distribution.
Unlike prior models that consider entanglement generation or quantum
memory coherence in isolation, we investigate the interplay between
path assignment strategies, coherence time constraints, and quantum
error correction (QEC), and how they jointly impact end-to-end communication
fidelity. We analyze both unencoded and encoded transmissions over
aggregated network paths, quantifying the effects of resource allocation
on transmission success. Our findings show that fidelity cannot be
optimized independently of memory lifetimes, and that while QEC can
enhance performance under specific conditions, it also imposes additional
constraints depending on network topology and path-length asymmetries.
This work provides a foundation for developing QoS-aware quantum routing
protocols that balance fidelity, throughput, and memory utilization---key
considerations for near-term quantum repeater networks.}
\end{abstract}
\maketitle

\section{Introduction }

\textcolor{black}{Large-scale quantum networks aim to connect users
distributed across geographically distant locations by establishing
entangled quantum states between them \citep{QNET1,NVnetworks1,NVnetworks2}.
Entanglement swapping operations at intermediate nodes will enable
long-distance connections, ultimately contributing to the realization
of a global quantum internet \citep{Q_internet}. Such a network will
offer fundamentally new capabilities, enabling users to harness quantum
technologies that surpass the performance of classical systems. For
example, connecting spatially separated quantum processors across
the network could significantly enhance the computational power of
distributed quantum computers \citep{QC_google,QC_Bill,Qcomp1,Qcomp2,Quantum_comp2,Quantum_comp3,Quantum_comp4}.
Secure communication between remote users will also be made possible
through quantum key distribution and related protocols \citep{QC_Bill,QKD03,QKD1,QKD2000,QKD_transmission}.
Moreover, linking multiple quantum sensors or clocks across nodes
can boost the precision and accuracy of quantum metrology applications,
such as clock synchronization and remote sensing \citep{QImaging2,QImaging3,QSensing}.}

\textcolor{black}{Despite the profound differences between classical
and quantum networks, many similarities remain---especially given
that quantum networks will be deployed atop existing telecommunication
infrastructure. In both cases, the number of physical resources available
at each node is inherently limited, leading to potential congestion.
This congestion can severely affect network performance by reducing
data throughput or degrading the fidelity of quantum states.}

\textcolor{black}{In classical networks, congestion and resource contention
are managed using a suite of traffic control techniques that collectively
define Quality of Service (QoS) \citep{q_capacity1,q_capacity2,q_capacity3}.
QoS metrics include bandwidth (data transfer rate), loss (fraction
of lost data), delay (end-to-end transmission time), and jitter (variability
in packet arrival times). These metrics guide network design and optimization
to ensure that performance requirements are met even under constrained
resources.}

\textcolor{black}{Similar considerations will be critical for the
design of quantum networks. However, the extension of QoS concepts
to the quantum domain must take into account fundamental physical
differences. For instance, in a quantum network, delays often require
temporary storage of quantum states in quantum memories, which are
inherently susceptible to decoherence. This leads to fidelity loss---an
effect absent in classical communication. As such, a quantum version
of QoS must incorporate both conventional performance metrics and
quantum-specific considerations such as memory coherence, noise, and
quantum error correction (QEC) overhead.}

\textcolor{black}{Recent efforts have begun to formalize QoS in the
quantum setting \citep{Wehener_QoS}. For example, Cicconetti et al.
\citep{Cicco} explored how trade-offs between fidelity, latency,
and throughput arise in realistic quantum architectures. Their work
showed that multipath transmission can enhance network resource utilization,
motivating our investigation of aggregated quantum networks (AQNs),
in which quantum information is distributed over multiple paths to
improve performance. In this work, we build upon that direction by
introducing a QoS-aware channel assignment protocol for AQNs in a
non-entanglement-based transmission model \citep{QEC2}. Unlike most
prior studies that focus on entanglement distribution, swapping, or
purification \citep{QROUTING1,QROUTING2,QROUTING3,Cicco}, our model
is based on the direct transmission of quantum information---encoded
or unencoded---through noisy channels. In this scenario, quantum
error correction may be applied to protect the transmitted state from
loss and noise \citep{redundancy_code,surfacecode,QEC1,GKP1,QEC2,QEC3,QEC4,QEC5}.
Interestingly, our analysis reveals that QEC does not always improve
fidelity. In regimes with short coherence times or unbalanced path
lengths, the overhead introduced by encoding can actually degrade
the overall performance. This nontrivial result highlights the importance
of developing integrated routing and encoding strategies that explicitly
account for the physical limitations of hardware.}

\textcolor{black}{Here, we focus on key features of a quantum network
that could pave the way for future in-depth investigations. Specifically,
we consider a setting where two distant nodes, each hosting multiple
users and a finite number of communication channels, aim to exchange
quantum information with a target fidelity. While selecting fidelity
as the primary figure of merit may appear restrictive, high-fidelity
transmission is essential for realizing the advantages of quantum
communication over classical methods. We model these two nodes as
connected by multiple paths of varying lengths, each consisting of
a series of lossy channels. To mitigate the effects of loss, we consider
encoding the quantum state into a higher-dimensional system using
QEC, distributing the encoded components across the available paths.}

\textcolor{black}{In AQNs, quantum packets (qubits or qudits) representing
the encoded state are sent through different paths and later recombined
at the receiver node for decoding. Thus, optimal channel assignment
plays a crucial role in maintaining fidelity. We propose a routing
protocol that distributes the encoded state across multiple channels
in response to user demands, guided by QoS-inspired metrics tailored
to the quantum setting. This framework allows us to define and optimize
QoS parameters for AQNs, extending the classical concept into the
quantum domain.}

\textcolor{black}{Importantly, our model departs from entanglement-based
approaches. No entanglement generation, swapping, or purification
steps are involved. Instead, we focus entirely on direct transmission
of quantum information through noisy channels, a relevant model for
near-term quantum networks constrained by hardware limitations.}

\textcolor{black}{This paper is organized as follows. In Sec. II,
we define QoS metrics adapted to aggregated quantum networks. In Sec.
III, we introduce a quantum router protocol for channel assignment
based on these metrics. In Sec. IV, we present a case study illustrating
how routing decisions impact fidelity under temporal delays and path
asymmetries. Finally, we conclude in Sec. V.}

\section{Quality of Service of an aggregated quantum network}

\begin{figure}[h]
\begin{centering}
\textcolor{black}{\includegraphics[scale=0.15]{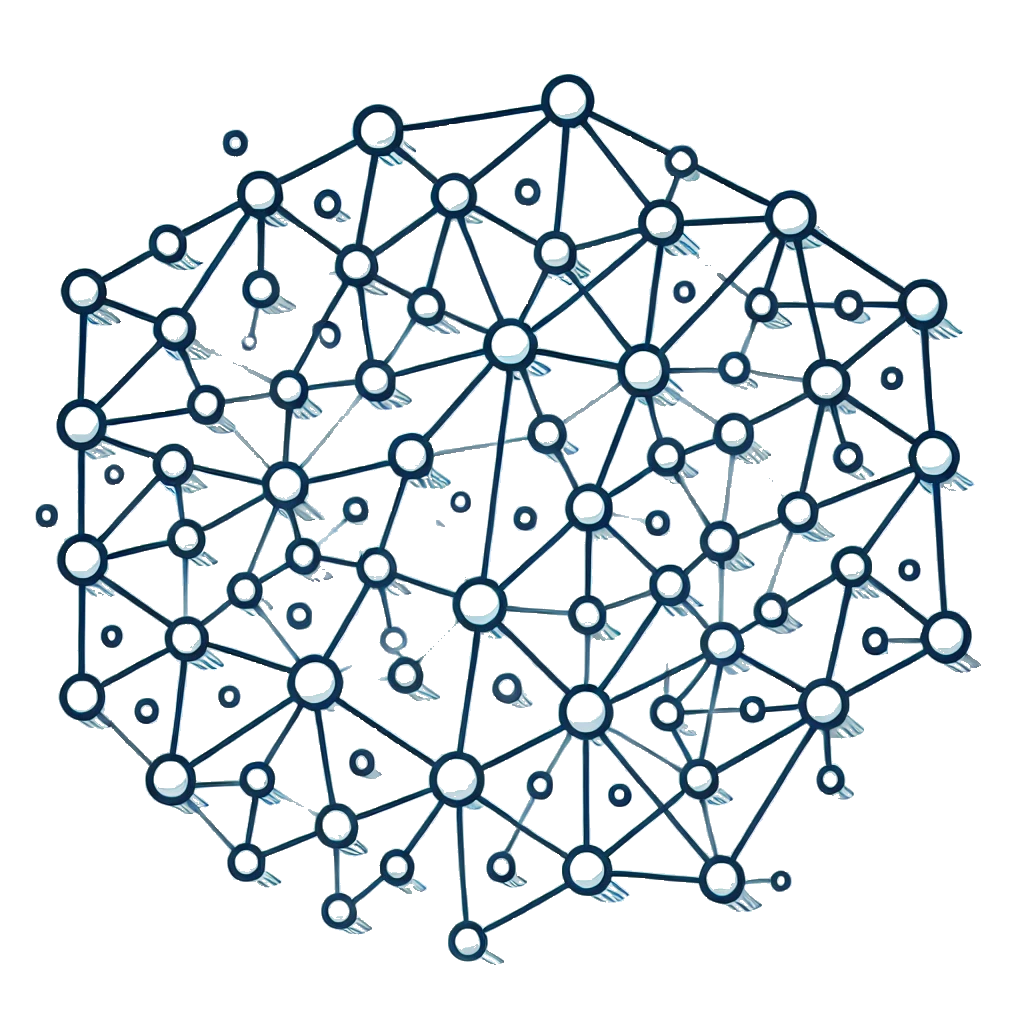}}
\par\end{centering}
\textcolor{black}{\caption{\label{fig:A-generic-quantum}A generic quantum network represented
by a set of vertices (nodes) connected by edges (links). Each link
can contain multiple channels. The set of links that connects two
nodes is a path.}
}
\end{figure}

\begin{figure}[h]
\begin{centering}
\includegraphics[scale=0.45]{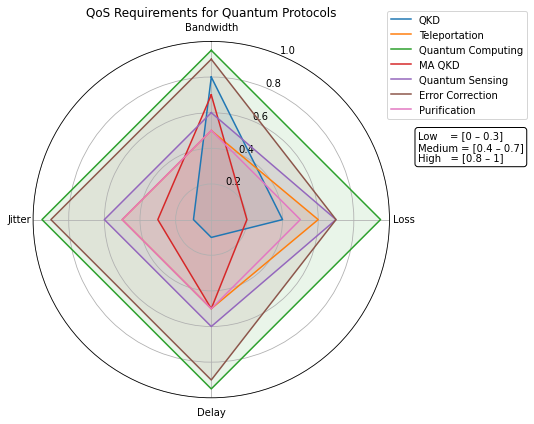}
\par\end{centering}
\caption{\label{fig:Radar-plot-of}The chart provides a qualitative comparison
of the performance requirements of various applications across key
QoS metrics. The normalized values are not drawn from experimental
data but are instead intended to illustrate conceptual differences
among protocols, helping guide the design of routing strategies suited
to specific use cases. The range of normalized values for ``low'',
``medium'' and ``high'' requirements is in the box below the legend. }

\end{figure}

\textcolor{black}{We now extend the QoS concepts introduced earlier
to the context of an Aggregated Quantum Network (AQN), beginning with
some basic definitions. It is important to emphasize that our model
does not rely on entanglement-based primitives. As a result, phenomena
such as stochastic link establishment arising from probabilistic entanglement
swapping---which are central to repeater-based architectures---are
not relevant in our setting. Instead, we consider a direct transmission
model in which quantum states are sent over lossy channels characterized
by fixed or known delays. In general, the network can be represented
as a graph, where nodes (vertices) are connected by communication
links (edges), as illustrated in Fig. \ref{fig:A-generic-quantum}.
}
\textcolor{black}{{} We will refer to the vertices of a quantum network
as $nodes$ and to the edges connecting them as $links$. A $link$
represents either a physical or logical connection between two nodes,
while a $channel$ denotes the specific transmission medium---physical
or logical---used to carry quantum information. More precisely, a
channel may correspond to a dedicated physical medium, such as an
optical fiber or free-space optical connection, or to a logical subdivision
within a multiplexed medium. Since a link may support multiple channels,
we define the $capacity$ of a link as the aggregate capacity of all
channels associated with it. A $path$ is the sequence of links and
intermediate devices that a quantum packet traverses en route from
source to destination. The length of a path, denoted by L, is the
sum of the lengths of its constituent links and is measured in kilometers.
In AQNs, quantum information is distributed across multiple parallel
paths---each potentially of different length---that collectively
connect a sender to a receiver. This set of paths constitutes an aggregated
path. In such a scenario, several QoS metrics must be redefined or
adapted to account for the unique features of AQNs.}
\begin{itemize}
\item \textcolor{black}{Bandwidth: the bandwidth of a single path is determined
by the capacity of the bottleneck link---i.e., the link along the
path with the lowest capacity. For an aggregated path, the bandwidth
is defined as the sum of the capacities of all constituent paths.
Bandwidth thus quantifies the maximum amount of information that can
be transmitted across the path per unit time, providing users with
an estimate of the available data rate.}
\item Loss: the loss of information transmitted across noisy channels can
be measured in several ways. We adopt the infidelity of the transmitted
state, $1-F$, where $F$ is the fidelity of the state received and
processed at the receiver's node. The infidelity might depend on several
errors, such as the channel loss, \textcolor{black}{parametrized by
the transmission coefficient $p=\eta e^{-L/L_{att}}$, where $L_{att}$
is the attenuation length that depends on the type of channel in use,
and $\eta$ is a loss term due to interactions with other components
of the network, such as quantum memories (QMs) and frequency converters.}
In an aggregated quantum network the infidelity depends also on the
decoherence time of the QMs used to store the quantum states at the
receiver's node \citep{temporal_delay}. 
\item Delay: the $delay$, $\tau_{P},$ of a path is defined as the time
interval between the time the quantum data packets are sent by the
sender node and the time they arrive at the receiver node. It is given
by $\tau_{P}=L/c+t_{c},$ where $L$ is the length of the path, $c$
is the speed of light traveling in the medium (optical fiber or free
space) while $t_{c}$ is the congestion time. The latter is the extra
time due to the waiting time some packets might require at the intermediate
nodes and the processing time. In fact, one can think that when the
physical resources travel across multiple nodes before reaching the
far end, some of those intermediate nodes might be not be immediately
accessible or could take some time to process the quantum data packet.
Therefore, some delay line must be added before the resources can
pass through that node. This time can be then considered as the sum
of all those extra delays. This definition of delay can be generalized
by including more temporal terms due to the queuing times. \textcolor{black}{The
congestion time is, therefore, dynamic and dependent on queuing and
processing times. \citep{reference_note1}.}
\item Jitter: this is defined as the time difference of the arrival times
of the information packets. Whereas in classical networks such a time
difference will cause a delay in the information flow, in \textcolor{black}{aggregated
quantum networks that information must be stored into QMs. Therefore,
the stored states will undergo a depolarization/dephasing pro}cess
that will deteriorate the fidelity of the final state, resulting in
a loss of information. Here we have assumed the jitter is negligible.
However, in a real implementation the jitter can be different from
zero and, in this case, must be considered.
\end{itemize}
\noindent \textcolor{black}{In the quantum networking regime, these
metrics can overlap. For instance in some models, particularly those
involving entanglement distribution or application-specific fidelity
guarantees, it may be useful to define an \textquoteleft effective
bandwidth\textquoteright{} that combines channel capacity with a minimum
required fidelity. However, in our work, we do not impose such constraints.
Our model evaluates the actual fidelity achieved under different channel
assignments but does not enforce a fidelity threshold. As a result,
bandwidth and loss are treated as separate performance metrics in
the unencoded case. In the encoded case, they may become indirectly
related via the redundancy of the quantum error correction code, but
we do not introduce an explicit effective bandwidth definition. Based
on the above quantum network definitions of QoS, we can now qualitatively
estimate the requirements of a few quantum protocols, }as shown in
the radar plot of Fig. \ref{fig:Radar-plot-of}. In a conventional
quantum key distribution (QKD) protocol the rate at which two users
can share a secret key is the main figure of merit. Therefore, the
bandwidth plays a fundamental role and it must be high \citep{QKD_adv}.
Similarly, for distributing quantum computing high bandwidth is necessary
to connect multiple nodes faster \textcolor{black}{\citep{QComp_Bill}.}
For memory assisted (MA) QKD and conventional purification protocols\textcolor{black}{{}
the bandwidth can be slightly lower, as quantum memories store the
quantum states increasing the probability of successfully sharing
information \citep{memory_ass}. Also, for quantum remote sensing,
we need to generate complex states like GHZ states, which need higher
quality (fidelity) \citep{QImaging4}. A lower bandwidth is required
for quantum teleportation since a single entangled state is sufficient
for its realization. Then, a small delay and the jitter for QKD protocols
do not affect the efficiency of these systems since a delayed secret
key and an irregular transmission of the arriving packets do not compromise
the security of the shared key. A medium requirement on both the delay
and jitter is, on the other hand, needed for purification, teleportation
and remote sensing as storing entangled states might affect the quality
of the final states due to decoherence processes in the QMs. Distributing
quantum computing is highly affected by delay and jitter as it is
fundamental that the packets arrive regularly and at precise times
at each node to be processed correctly. Finally, the loss of information
has a low impact on a MA-QKD protocol because the states are stored
into QMs before extracting a secret key. This loss has a medium effect
on conventional QKD systems as it might reduce the secret key rate.
Analogously for purification the loss of information has a medium
impact on the efficiency of this protocol because several entangled
states need to be created to increase the fidelity of the final state.
Then, for distributing quantum computation the quantum error correction
codes in use can restore the loss of information with some probability,
having, thus, a medium impact on these systems. On the contrary, losses
have a more detrimental effect on the sensitivity of a sensing protocol
and can move us away from the Heisenberg limit \citep{Heisenberg_Lim}.
}This is especially true as the kind of resource states we are generating
can be completely destroying if a single qubit it loss.

\textcolor{black}{Figure \ref{fig:Radar-plot-of} presents a radar
chart summarizing the QoS requirements of various quantum applications.
These requirements exhibit diverse performance profiles: some applications
demand high fidelity, while others are more tolerant of latency or
jitter. Notably, the chart illustrates that no single routing policy
can optimally satisfy all applications simultaneously---particularly
in networks with limited bandwidth or highly variable channel quality
across different paths. This observation raises a fundamental design
challenge for near-term quantum networks: how should a router allocate
transmission resources---such as quantum channels along multiple
paths---to meet the potentially conflicting QoS demands of multiple
users? In the following section, we address this question by introducing
a time-slotted quantum router model and analyzing several channel
assignment strategies under resource-constrained conditions.}

\section{The routing protocol}

\textcolor{black}{We consider a scenario in which multiple users issue
requests to transmit quantum information---using either qubits or
qudits---over a quantum network composed of multiple transmission
paths and coordinated by a quantum router. The router is responsible
for assigning available quantum channels to these requests in a manner
that satisfies or balances relevant Quality of Service (QoS) metrics.
Our model adopts a time-slotted structure, in which time is divided
into discrete, fixed-duration intervals (slots). Each slot corresponds
to the transmission window of a quantum \textquotedblleft packet,\textquotedblright{}
defined here as a group of qudits sent from a user to a destination
node.}

\textcolor{black}{All slots are assumed to have equal duration, determined
by the time required to transmit a single packet (i.e., a batch of
qudits). We do not assume global synchronization across the entire
network, as this would require impractically precise timing mechanisms.
Instead, we assume local synchronization between directly connected
nodes, which is consistent with the capabilities of current quantum
key distribution (QKD) systems, where local time alignment can be
achieved via classical signaling or embedded timing protocols \citep{sync1,sync2}.}

\textcolor{black}{The quantum router does not directly transmit quantum
information but acts as a decision-making component that allocates
quantum channels and manages resource assignment across the network.
It interfaces with underlying quantum switches, which are responsible
for the physical transmission of qudits. The router performs higher-level
control functions, including receiving user requests (e.g., the number
of qudits to be sent), querying or maintaining the current state of
each transmission path (such as channel availability and estimated
loss or delay), and assigning channels to users according to a specified
routing policy.}

\textcolor{black}{By contrast, quantum switches operate at the physical
layer, forwarding qudits based on the router\textquoteright s assignments.
Routers and switches exchange control information over a classical
control channel, in line with standard practices in quantum network
control architectures \citep{Qcomp_Rod}. Users communicate with the
quantum router via a classical interface, specifying the destination
node, the number of qudits to transmit, and optionally a QoS preference.}

\textcolor{black}{In our model, we assume that the quantum router
has knowledge of the following:}
\begin{itemize}
\item \textcolor{black}{The number of available channels along each path, }
\item \textcolor{black}{The estimated delay and length of each path (which
affect fidelity and timing),}
\item \textcolor{black}{The coherence time of the quantum memories at the
destination node, used to determine whether the qudits will survive
the full transmission time, }
\item \textcolor{black}{The set of current user requests. }
\end{itemize}
\textcolor{black}{The router does not perform entanglement distribution,
purification, or entanglement swapping. Instead, our model focuses
on the direct transmission of quantum information, where quantum memories
serve primarily as buffers to temporarily store qudits until transmission
can be completed. This architectural choice enables the router to
reject or reroute requests whose expected delay exceeds the coherence
time of the available memory, as illustrated in the examples presented
in Sec. IV.}

\subsection{\textcolor{black}{Routing protocol description}}

\begin{figure}[h]
\begin{centering}
\textcolor{black}{\includegraphics[scale=0.25]{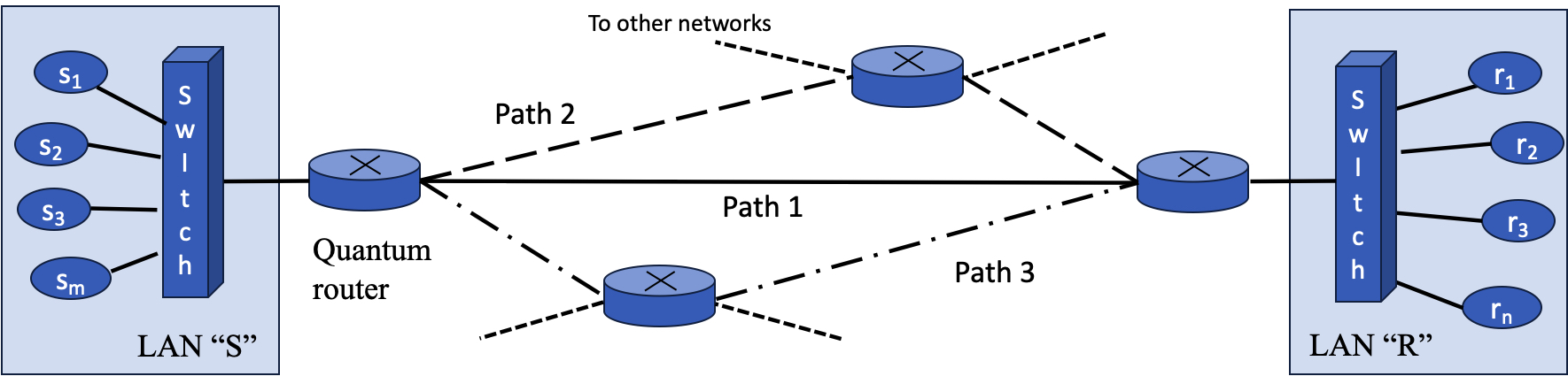}}
\par\end{centering}
\textcolor{black}{\caption{\label{fig:The-routing-protocol.}\textcolor{black}{Schematic illustration
of a wide-area quantum network (WAN) with quantum routers. Two local
area networks (LANs), labeled \textquotedblleft S\textquotedblright{}
(source) and \textquotedblleft R\textquotedblright{} (receiver), are
connected via three distinct paths: a solid line (path 1), dashed
lines (path 2), and dash-dotted lines (path 3). These paths differ
in length and may each contain an arbitrary number of quantum channels,
over which information can be transmitted from LAN S to LAN R. A quantum
router assigns available channels for transmitting quantum packets
according to a specified operational policy based on user requests,
or it may deny requests if network resources are insufficient to meet
the stated requirements. Each path may also include intermediate routers
that connect to other networks.}}
}
\end{figure}

\textcolor{black}{Let us now describe our routing protocol for a simple
wide area network (WAN) as illustrated in Fig. \ref{fig:The-routing-protocol.}.
}
\textcolor{black}{{} We consider two local area networks (LANs), denoted
LAN \textquotedblleft S\textquotedblright{} (source) and LAN \textquotedblleft R\textquotedblright{}
(receiver), containing m and n users, respectively. Both LANs are
connected to quantum routers, which are, in turn, linked via multiple
paths of varying delay. Each path may include an arbitrary number
of quantum channels over which information is transmitted. While a
real-world quantum network may support many such paths between distant
LANs, for illustrative purposes Fig. 3 depicts three representative
paths: a solid line (path 1), dashed lines (path 2), and dash-dotted
lines (path 3).}

\textcolor{black}{The objective is to allocate the available resources
in LAN \textquotedblleft S\textquotedblright{} across the various
channels connecting it to LAN \textquotedblleft R.\textquotedblright{}
To this end, we envision a routing mechanism that follows a multi-step
protocol. Initially, users in LAN \textquotedblleft S\textquotedblright{}
issue requests to communicate with users in LAN \textquotedblleft R,\textquotedblright{}
specifying desired performance criteria---such as a minimum threshold
fidelity for received quantum information or a target transmission
rate (request step). The quantum router then evaluates whether the
available resources in both LANs are sufficient to fulfill these requirements
(processing step).}

\textcolor{black}{If a request cannot be satisfied with the current
resource state, the router temporarily queues it for a fixed number
of time steps. If the network lacks the capacity to meet the request
even under future conditions, the router issues a denial signal, notifying
the user that the communication cannot proceed and awaits the next
input. Otherwise, the router proceeds to the assignment step, in which
it distributes available resources across the selected channels. We
now examine each of these steps in more detail.}

\subsection{\textcolor{black}{Request step}}

\textcolor{black}{In the request step, the primary figure of merit
is evaluated. Upon receiving an input request, the quantum router
first determines whether error protection is required to guarantee
the desired fidelity (if that is the desired metric) of the transmitted
quantum state. If error protection is necessary, the router then identifies
the appropriate operational regime. Multiple regimes may be considered
depending on user requirements. For example, the router might prioritize
a single user or a small subset of users, thereby limiting access
to the transmission paths for the majority. Alternatively, it may
ensure that most users gain access to the available paths. While numerous
scenarios are possible, in the following section we focus on three
representative regimes, leaving a more comprehensive analysis for
future work.}

\textcolor{black}{Once the request step is completed, the router proceeds
to the processing step. Here, it analyzes the available resources
in LAN \textquotedblleft S,\textquotedblright{} the coherence times
of quantum memories in LAN \textquotedblleft R,\textquotedblright{}
and the characteristics of the connecting paths. Based on this information
and the selected operational regime, the router assigns channels to
the users in LAN \textquotedblleft S.\textquotedblright{} This step
is critical for optimizing protocol performance, and we provide a
simple illustrative example in the next section.}

\textcolor{black}{Finally, after processing, the quantum router instructs
the switch device to establish connections for users to transmit their
data. The router then resets and initializes for the next operational
cycle. }

\section{Routing assignment in the unencoded scenario}

\textcolor{black}{Selecting an appropriate channel assignment can
enhance network performance. Quality of Service (QoS) features inform
this selection; here, we focus on two key aspects: bandwidth and loss.
We assume that channels have limited capacity and that information
may be lost during transmission or degraded when stored in quantum
memories (QMs). In this section, we present an example illustrating
how a router can assign multiple channels between two distant nodes,
S and R, each hosting a variable number of users seeking to exchange
information. We assume that S and R are connected by $N_{c}$ channels
with $N_{c}/2$ channels contained into a path with delay $\tau_{1}$
and the remaining $N_{c}/2$ channels in a path with delay $\tau_{2}>\tau_{1}.$
Let us further assume that a router at node S assigns channels to
users, across which they transmit their quantum states to node R.
A natural question then arises: what is the optimal channel assignment?
To answer this question we can explore three possible scenarios. In
scenario (a), called the $greedy$ $regime$, a single user maximizes
the quality (fidelity) of their transmitted state. In scenario (b),
the goal is to satisfy the requirements of as many users as possible
($balanced$ $regime$). In scenario (c), a limited number of users
aim to maximize their fidelities, subject to the condition that the
differences among them remain small ($restricted$ $regime$). Depending
on the selected regime the router will assign the available channels
accordingly. The question now is: what is the optimal assignment strategy
in each of the three regimes? In this Section we show with an example
what is the best assignment when users send information using five
packets. Consider that $N_{c}=10$ and that the capacity of each channel
can transmit a qudit of dimension 9 per time unit. A user could send
a single lower-dimensional qudit or, alternatively, two qudits of
dimension 3 (qutrits). For simplicity, let us assume that $\eta_{1}=\eta_{2}=1$
and $t_{c_{1}}=t_{c_{2}}=0.$ Table \ref{tab:unencoded tab} presents
all the possible assignments of the channels of path 1 and path 2
among $N_{u}$ users, with $N_{u}$ being the highest number of users
that can be served given the above conditions. For instance, the first
line of table \ref{tab:unencoded tab} gives the number of qudits
traveling in each path for two assignments given to two users. In
this case, under assignment 1, one user transmits a packet of five
qudits of dimension 7 through the channels along path 1, while another
user sends their five qudits of the same dimension through the channels
of path 2. This configuration corresponds to the greedy regime, in
which the router assigns the lowest-loss channels to a single user,
leaving the remaining channels (in path 2) to the other user. Consequently,
the fidelity of the \textquotedbl greedy\textquotedbl{} user reaches
the maximum achievable with the available resources, assuming $p_{1}$
= 0.95 and no decoherence, as shown by the solid line in Fig. 4. In
contrast, the fidelity of the packets sent by the other user is the
lowest, as indicated by the cross symbols in Fig. 4.}
\begin{table}
\begin{centering}
\textcolor{black}{}%
\begin{tabular}{|c|c|c|}
\cline{1-2} \cline{2-2} 
\textcolor{black}{Assignment 1} & \textcolor{black}{Assignment 2} & \multicolumn{1}{c}{}\tabularnewline
\hline 
\textcolor{black}{}%
\begin{tabular}{|c|c|}
\hline 
\textcolor{black}{path 1} & \textcolor{black}{path 2}\tabularnewline
\hline 
\textcolor{black}{5} & \textcolor{black}{0}\tabularnewline
\hline 
\textcolor{blue}{4} & \textcolor{blue}{1}\tabularnewline
\hline 
\textcolor{blue}{3} & \textcolor{blue}{2}\tabularnewline
\hline 
\textcolor{black}{5} & \textcolor{black}{5}\tabularnewline
\hline 
\end{tabular} & \textcolor{black}{}%
\begin{tabular}{|c|c|}
\hline 
\textcolor{black}{path 1} & \textcolor{black}{path 2}\tabularnewline
\hline 
\textcolor{black}{0} & \textcolor{black}{5}\tabularnewline
\hline 
\textcolor{blue}{1} & \textcolor{blue}{4}\tabularnewline
\hline 
\textcolor{blue}{2} & \textcolor{blue}{3}\tabularnewline
\hline 
\textcolor{black}{5} & \textcolor{black}{5}\tabularnewline
\hline 
\end{tabular} & \textcolor{black}{}%
\begin{tabular}{|c|}
\hline 
\textcolor{black}{$N_{u}$}\tabularnewline
\hline 
\textcolor{black}{2}\tabularnewline
\hline 
\textcolor{black}{2}\tabularnewline
\hline 
\textcolor{black}{2}\tabularnewline
\hline 
\textcolor{black}{4}\tabularnewline
\hline 
\end{tabular}\tabularnewline
\hline 
\end{tabular}
\par\end{centering}
\textcolor{black}{\caption{\label{tab:unencoded tab}\textcolor{black}{Assignments of the number
of qudits transmitted along two paths in the unencoded scenario. Each
row corresponds to a possible channel assignment among $N_{u}$ users.
Configurations highlighted in blue are affected by decoherence occurring
in the quantum memories (QMs) located at node R. These QMs store packets
that arrive earlier, holding the quantum states until the packets
traveling through the longer-delay path are ready to be processed.}}
}
\end{table}
\textcolor{black}{{} Next, in the restricted regime, the router assigns
channels to users with the goal of minimizing the difference between
their fidelities. This condition is satisfied by the configurations
represented by the dotted and dash-dotted lines in Fig. 4. The balanced
regime, in this case, corresponds to one of the three configurations
listed in Table I, since the maximum number of users that can utilize
the network is two. The mathematical expressions for the fidelities
shown in Fig. 4 are provided in Appendix A.}

\begin{figure}
\begin{centering}
\includegraphics[scale=0.22]{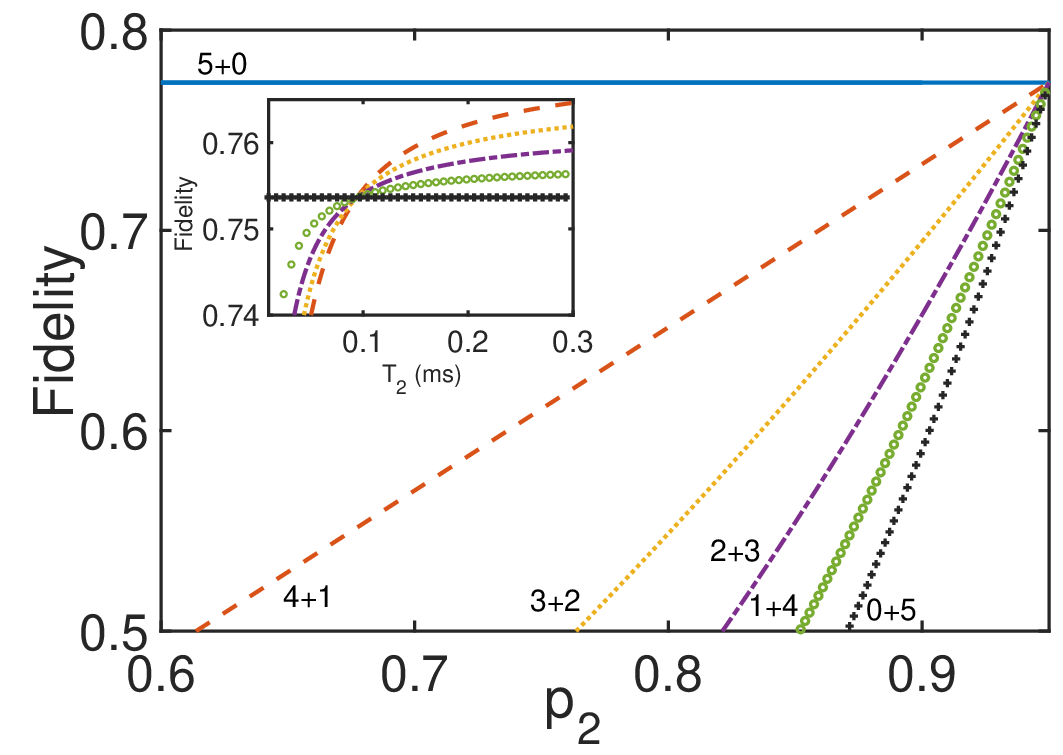}
\par\end{centering}
\caption{\label{fig:Fidelities-of-the-unencoded}Fidelities of the received
packets at the receiver node as a function of the transmission probability
of path 2, $p_{2},$ corresponding to qudit assignments listed in
Table I with $p_{1}=0.9$ and $T_{2}\rightarrow\infty$. The label
of each curve refers to the number of qudits sent through path 1 and
path 2, corresponding to the assignments of Table \ref{tab:unencoded tab}.
The inset depicts the fidelities of the packets at the receiver node
versus the coherence time, $T_{2},$ for $p_{1}=0.95$ and $p_{2}=0.$945.}
\end{figure}

\textcolor{black}{It is now interesting to see the effects of a finite
coherence time in such a scenario. The configurations, highlighted
in blue in Table \ref{tab:unencoded tab}, require quantum memories
at the receiver's node due to the fact that the paths have different
delay (length in this case) and, therefore, the receiver node must
store the quantum states arriving earlier before the total packet
can be processed. However, at low coherence times, the decoherence
might affect a configuration with the majority of the qudits traveling
in the shorter path much more than a configuration with a larger number
of qudits traveling in the longer path, as shown in the inset of Fig.
\ref{fig:Fidelities-of-the-unencoded}. One can see that at $T_{2}<0.1$
ms and at $p_{2}=0.945$ the fidelity of the transmitted packet corresponding
to the assignment 4+1 (3+2) is higher than the fidelity corresponding
to the assignment 1+4 (2+3). For these configurations it is then fundamental
that the router is informed with the coherence time of the QMs at
the receiver node to optimally assign the channels to the users. Remarkably,
the inset shows that the fidelities of those configurations all cross
with the 0+5 configuration at $T_{2}^{\dagger}=0.1$ ms. This means
that for $T_{2}<T_{2}^{\dagger}$ the aggregation scenario is not
convenient for a possible router's assignment. Finally, the regimes
defined above are not affected by finite coherence times. }

\section{Routing assignment in the encoded scenario}

\textcolor{black}{In this section, we illustrate, via an example,
how a router assigns the channels between two nodes by selecting between
a conventional network where the information is encoded and transmitted
across one or multiple paths with different delays. For those different
scenarios we compare the fidelities with the analytical formulas obtained
in \citep{temporal_delay}. We denote with ``$F_{i+j}"$ the fidelity
of the configuration $i+j$ in which $i$ $(j)$ qudits have been
transmitted across the first (second) path.}

Let us consider two distant nodes, S and R, exchanging information
using the quantum Reed-Solomon (QRS) code \citep{Reed_Sal} in a quantum
aggregation scenario. Similarly to the unencoded case, we assume here
that the number of users in S is a variable and node S and R are connected
by path 1 and path 2. For a fixed transmission probability $p_{1}$
of the channels in path 1, the best assignment depends on the transmission
probability of the channels in path 2 and on the coherence time of
the QMs at node R. Let us illustrate the answer to this question with
a simple example. Consider that $N_{c}=10$ and the capacity of each
channel can transmit a qudit of dimension 9 per time unit. For simplicity,
we also assume that $\eta_{1}=\eta_{2}=1$ and $t_{c_{1}}=t_{c_{2}}=0.$
\textcolor{black}{Now, each row of Table \ref{tab:emper_reg} contains
the number of qudits that can be sent per path with the above conditions
(highest capacity of the channel and number of channels per path).
For instance, in the first row of Assignment 1, a single user encodes
their information with a 7 dimensional QRS code, with 5(2) qudits
of dimension 7 traveling in path 1(2), respectively. Assignment 2
allows two users to encode their information with a 3 dimensional
QRS code by sending 3 qutrits across path 2. In this case, the highest
capacity have been achieved by the transmitted qudits therefore no
more assignments are possible (i.e., 0 qudits in Assignment 3). On
the contrary, the eighth row of Table \ref{tab:emper_reg} shows that,
when all users are encoding their state with a 3-dimensional QRS code,
three Assignments are possible. In Assignment 1 two users are sending
their three qutrits on path 1, two users send one qutrit on path 1
and two qutrits on path 2 in Assignment 2, and other two users send
three qutrits on path 2 in Assignment 3.}
\begin{table}
\begin{centering}
\textcolor{black}{}%
\begin{tabular}{|c|c|c|c|c|c|}
\cline{1-5} \cline{2-5} \cline{3-5} \cline{4-5} \cline{5-5} 
\textcolor{black}{Assignment 1} &  & \textcolor{black}{Assignment 2} &  & \textcolor{black}{Assignment 3} & \multicolumn{1}{c}{}\tabularnewline
\hline 
\textcolor{black}{}%
\begin{tabular}{|c|c|}
\hline 
\textcolor{black}{path 1} & \textcolor{black}{path 2}\tabularnewline
\hline 
\textcolor{blue}{5} & \textcolor{blue}{2}\tabularnewline
\hline 
\textcolor{blue}{4} & \textcolor{blue}{3}\tabularnewline
\hline 
\textcolor{blue}{3} & \textcolor{blue}{4}\tabularnewline
\hline 
\textcolor{blue}{2} & \textcolor{blue}{5}\tabularnewline
\hline 
\textcolor{black}{5} & \textcolor{black}{0}\tabularnewline
\hline 
\textcolor{blue}{4} & \textcolor{blue}{1}\tabularnewline
\hline 
\textcolor{blue}{3} & \textcolor{blue}{2}\tabularnewline
\hline 
\textcolor{black}{3} & \textcolor{black}{0}\tabularnewline
\hline 
\textcolor{black}{3} & \textcolor{black}{0}\tabularnewline
\hline 
\textcolor{blue}{2} & \textcolor{blue}{1}\tabularnewline
\hline 
\end{tabular} &  & \textcolor{black}{}%
\begin{tabular}{|c|c|}
\hline 
\textcolor{black}{path 1} & \textcolor{black}{path 2}\tabularnewline
\hline 
\textcolor{black}{0} & \textcolor{black}{3}\tabularnewline
\hline 
\textcolor{blue}{1} & \textcolor{blue}{2}\tabularnewline
\hline 
\textcolor{blue}{2} & \textcolor{blue}{1}\tabularnewline
\hline 
\textcolor{black}{3} & \textcolor{black}{0}\tabularnewline
\hline 
\textcolor{black}{0} & \textcolor{black}{5}\tabularnewline
\hline 
\textcolor{blue}{1} & \textcolor{blue}{4}\tabularnewline
\hline 
\textcolor{blue}{2} & \textcolor{blue}{3}\tabularnewline
\hline 
\textcolor{blue}{2} & \textcolor{blue}{1}\tabularnewline
\hline 
\textcolor{blue}{1} & \textcolor{blue}{2}\tabularnewline
\hline 
\textcolor{blue}{2} & \textcolor{blue}{1}\tabularnewline
\hline 
\end{tabular} &  & \textcolor{black}{}%
\begin{tabular}{|c|c|}
\hline 
\textcolor{black}{path 1} & \textcolor{black}{path 2}\tabularnewline
\hline 
\textcolor{black}{0} & \textcolor{black}{0}\tabularnewline
\hline 
\textcolor{black}{0} & \textcolor{black}{0}\tabularnewline
\hline 
\textcolor{black}{0} & \textcolor{black}{0}\tabularnewline
\hline 
\textcolor{black}{0} & \textcolor{black}{0}\tabularnewline
\hline 
\textcolor{black}{0} & \textcolor{black}{0}\tabularnewline
\hline 
\textcolor{black}{0} & \textcolor{black}{0}\tabularnewline
\hline 
\textcolor{black}{0} & \textcolor{black}{0}\tabularnewline
\hline 
\textcolor{black}{0} & \textcolor{black}{3}\tabularnewline
\hline 
\textcolor{black}{1} & \textcolor{black}{2}\tabularnewline
\hline 
\textcolor{blue}{1} & \textcolor{blue}{2}\tabularnewline
\hline 
\end{tabular} & \textcolor{black}{}%
\begin{tabular}{|c|}
\hline 
\textcolor{black}{$N_{u}$}\tabularnewline
\hline 
\textcolor{black}{3}\tabularnewline
\hline 
\textcolor{black}{3}\tabularnewline
\hline 
\textcolor{black}{3}\tabularnewline
\hline 
\textcolor{black}{3}\tabularnewline
\hline 
\textcolor{black}{2}\tabularnewline
\hline 
\textcolor{black}{2}\tabularnewline
\hline 
\textcolor{black}{2}\tabularnewline
\hline 
\textcolor{black}{6}\tabularnewline
\hline 
\textcolor{black}{6}\tabularnewline
\hline 
\textcolor{black}{6}\tabularnewline
\hline 
\end{tabular}\tabularnewline
\hline 
\end{tabular}
\par\end{centering}
\textcolor{black}{\caption{\label{tab:emper_reg}\textcolor{black}{Assignments of the number
of qudits traveling along two paths in the aggregation scenario. Each
row corresponds to a possible assignment of channels among $N{{}_u}$users.
Configurations affected by decoherence in the quantum memories (QMs)
located at node R are highlighted in blue. The QMs are needed when
the number of qudits arriving at R via the shorter path is not sufficient
to decode the state. The dimension of the qudits used is constrained
by the total capacity of each path.}}
}
\end{table}
\textcolor{black}{{} The last column of Table \ref{tab:emper_reg} contains
the highest number of users, $N_{u},$ that can be served given the
above conditions. We define the degeneracy of an assignment as the
highest number of users that can encode their state with the quantum
packets of that specific assignment. In the example above, for the
first row of Table \ref{tab:emper_reg} the degeneracy of Assignment
1 is one whereas the degeneracy of Assignment 2 is two as two users
can encode their state with a 3 dimensional QRS code. For simplicity,
for the assignment $i$ ($i\in\mathbb{N}^{*})$ with degeneracy higher
than one we refer to user $i$ as a single user that represents that
assignment. }

\begin{figure}[h]
\includegraphics[scale=0.2]{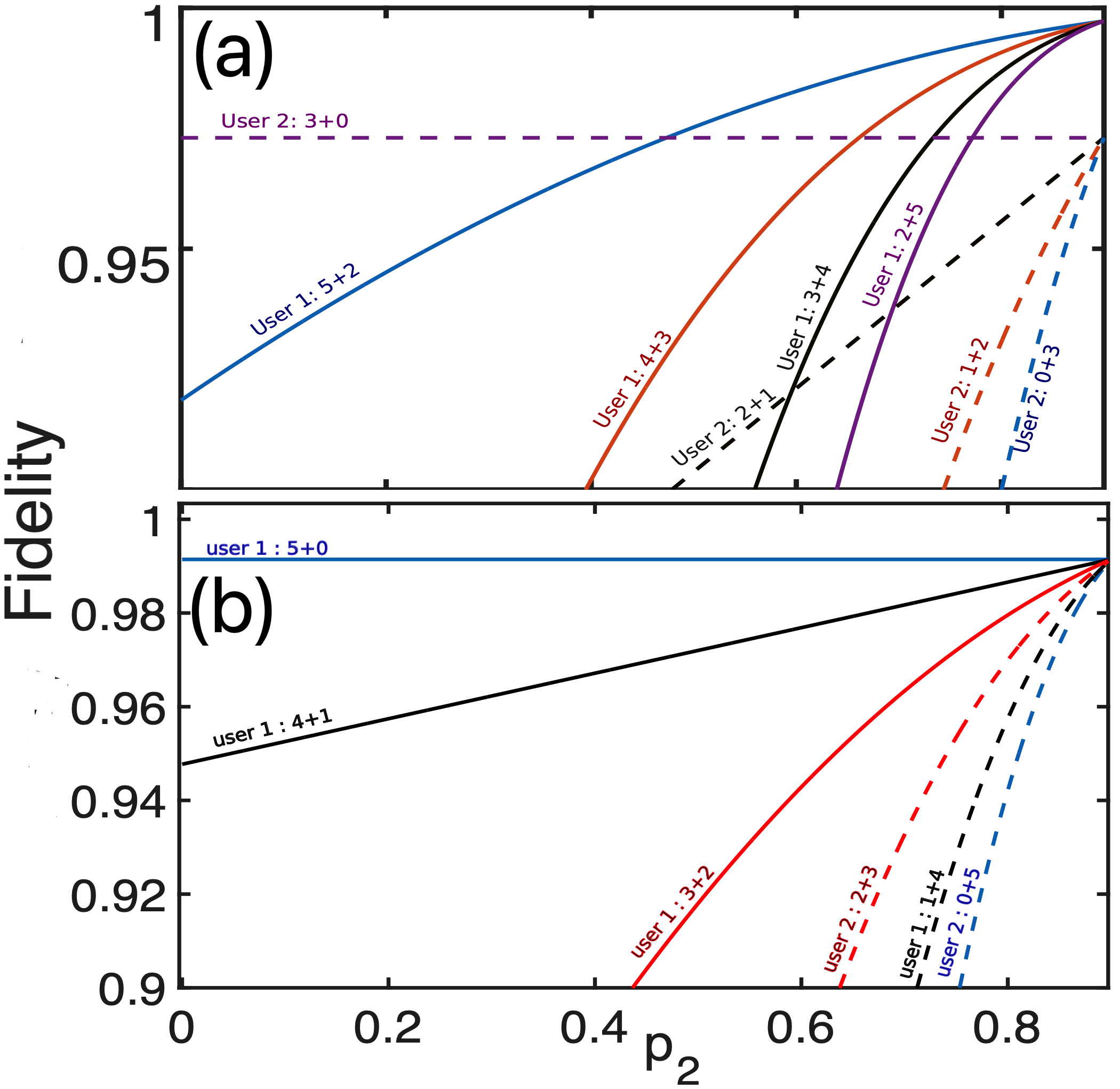}

\caption{\label{fig:(a)-and-(b)}\textcolor{black}{Fidelities of the decoded
states at the receiver node versus the transmission probability of
path 2, $p{{}_2}$, corresponding to the qudit assignments of Table
II, with $p{{}_1}$ = 0.9 and T\textsubscript{2}$ $\textrightarrow{}
\ensuremath{\infty}. Curves of the same color correspond to each row
in Table II. Solid (dashed) curves refer to user 1 (2), respectively.
The fidelities in (a) can be used to identify the greedy regime---since
some configurations employ the highest qudit dimension---the restricted
regime---because some fidelity curves intersect, thereby minimizing
their difference---and the balanced regime---due to configurations
involving the lowest number of qudits. Panel (b) shows fidelities
relevant for identifying both the restricted and greedy regimes. Specifically,
the restricted regime includes configurations with closely matching
fidelities, while the greedy regime includes configurations whose
performance is independent of the coherence time.}}
\end{figure}

\textcolor{black}{Now we analyze in more detail the various assignments
and the regimes with which they can be associated. First}ly, due to
the different lengths of the two paths some assignments (blue lines
in the table) might be affected by a temporal delay as shown in \citep{temporal_delay}.
This can significantly change the resulting fidelity depending on
the coherence time, $T_{2},$ of the quantum memories in use. Another
factor to consider is the length of path 2. Let us begin our analysis
with the asymptotic case $T_{2}\rightarrow\infty$ (ideal memories).
Figure \ref{fig:(a)-and-(b)}(a) 
 shows the fidelities of the transmitted states of user 1 and user
2, respectively, for the first four configurations of Table \ref{tab:emper_reg}.
Here the curves of the same colors correspond to the fidelities of
each single row and the solid (dashed) curves represent the fidelities
of the transmitted states of user 1 (2), respectively, for a transmission
coefficient of path 1 $p_{1}=0.9$. One can see that, when more qudits
are traveling in the shorter channel (red and blue curves),\textcolor{black}{{}
user 1 consistently achieves higher fidelities than user 2. In contrast,
when more qudits travel along path 2 (purple and yellow curves), the
advantage for user 1 occurs only for certain values of $p_{2}$ (crossing
points between curves of the same color). This behavior can be explained
by the fact that in the aggregation scenario the fidelities of the
configurations most packets travel through the longer channels strongly
depend on the length of that path. The crossing points of Fig. \ref{fig:(a)-and-(b)}(a)
can be used to determine the optimal distribution of quantum packets
in the restricted regime, where the goal is to minimize the fidelity
difference between users. }

\begin{figure}[h]
\begin{centering}
\includegraphics[scale=0.2]{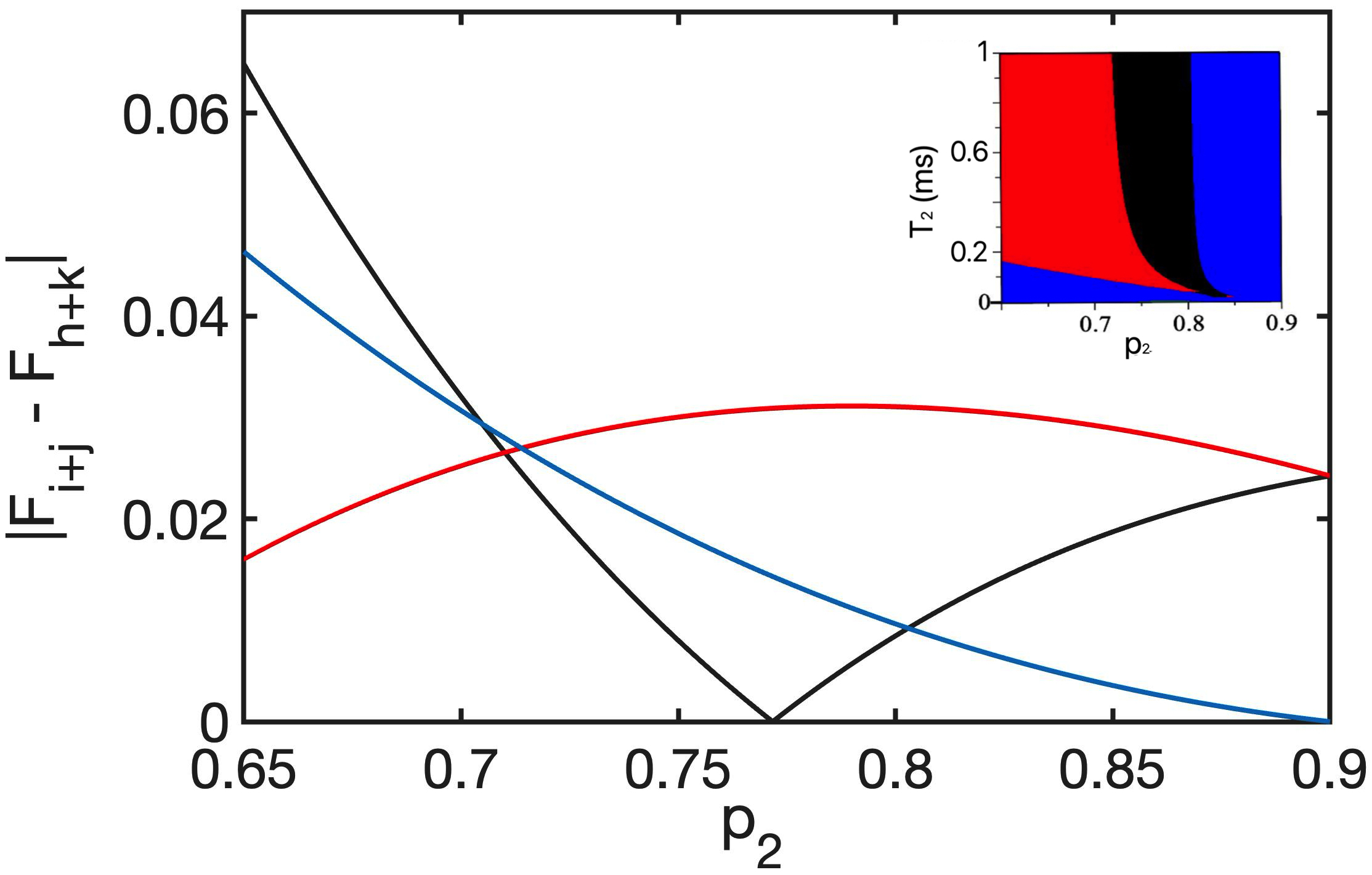}
\par\end{centering}
\caption{\label{fig:Difference-of-thefidelities}\textcolor{black}{Absolute
value of the difference between the fidelities of the transmitted
states, where $i,$ $j,$ $k$ and $l$ correspond to the following
configurations: 3+4 and 2+1 (red curve), 3+2 and 2+3 (blue curve),
and 2+5 and 3+0 (black curve). In the restricted regime, the router
assigns the 3+4 and 2+1 configuration (red curve) for 0.65 <$p_{2}$<
0.71, the 2+5 and 3+0 configuration (black curve) for 0.71 <$p_{2}<$
0.81, and the 4+3 and 1+2 configuration (blue curve) for $p_{2}>$
0.81. The inset shows the optimal configurations when fidelities are
affected by the coherence time $T_{2},$ by plotting $T_{2}$ versus
$p_{2}.$}}
\end{figure}

Next Fig. \ref{fig:(a)-and-(b)}(b) shows the fidelities of the transmitted
states for the corresponding configurations of Table \ref{tab:emper_reg}.
Here, the curves do not cross because both users send the same number
of qudits, although user 1 utilizes more channels in the shorter path
than user 2. The results in Fig. \ref{fig:(a)-and-(b)}(b) can be
used for the greedy regime by selecting the communication path with
the highest fidelity and for the restricted regime by analyzing the
plot of the difference of the fidelities. For simplicity, to determine
the optimal configuration for the restricted regime we plot in Figure
\ref{fig:Difference-of-thefidelities} 
only the difference of the fidelities of the transmitted states of
user 1 and user 2 of the 3+2 and 2+3 configuration (red curve), 4+3
and 1+2 configuration (blue curve) and 3+4 and 2+1 configuration (black
curve) versus $p_{2}$ . As one can see the graph can be divided into
three parts for certain range of $p_{2}.$ Specifically, for $0.65<p_{2}<0.71,$
the configurations that minimize the difference between the fidelities
of user 1 and user 2 correspond to distributing the resource according
to 3+2 the configuration for user 1 and 2+3 for user 2 (red curve).
Next, for $0.71<p_{2}<0.8,$ the black curve minimizes the difference
of the fidelities corresponding to the 3+4 configuration for user
1 and 2+1 for user 2. Finally at $p_{2}>0.8$ the smallest difference
of fidelities is given by the 4+3 configuration for user 1 and 1+2
configuration for user 2.

Further, the fidelities of the balanced regime corresponding correspond
to the communication paths in the last three rows of table \ref{tab:emper_reg}.
\textcolor{black}{Given the available resources, up to 6 users can
be served, as each assignment has a degeneracy of 2. Even for this
regime subcases may arise in which users might want to minimize the
difference in their fidelities (fair balanced regime) o}r assign most
of the resources traveling through the shorter channel to one user
(unfair balanced regime). 

\begin{figure}[h]
\begin{centering}
\includegraphics[scale=0.18]{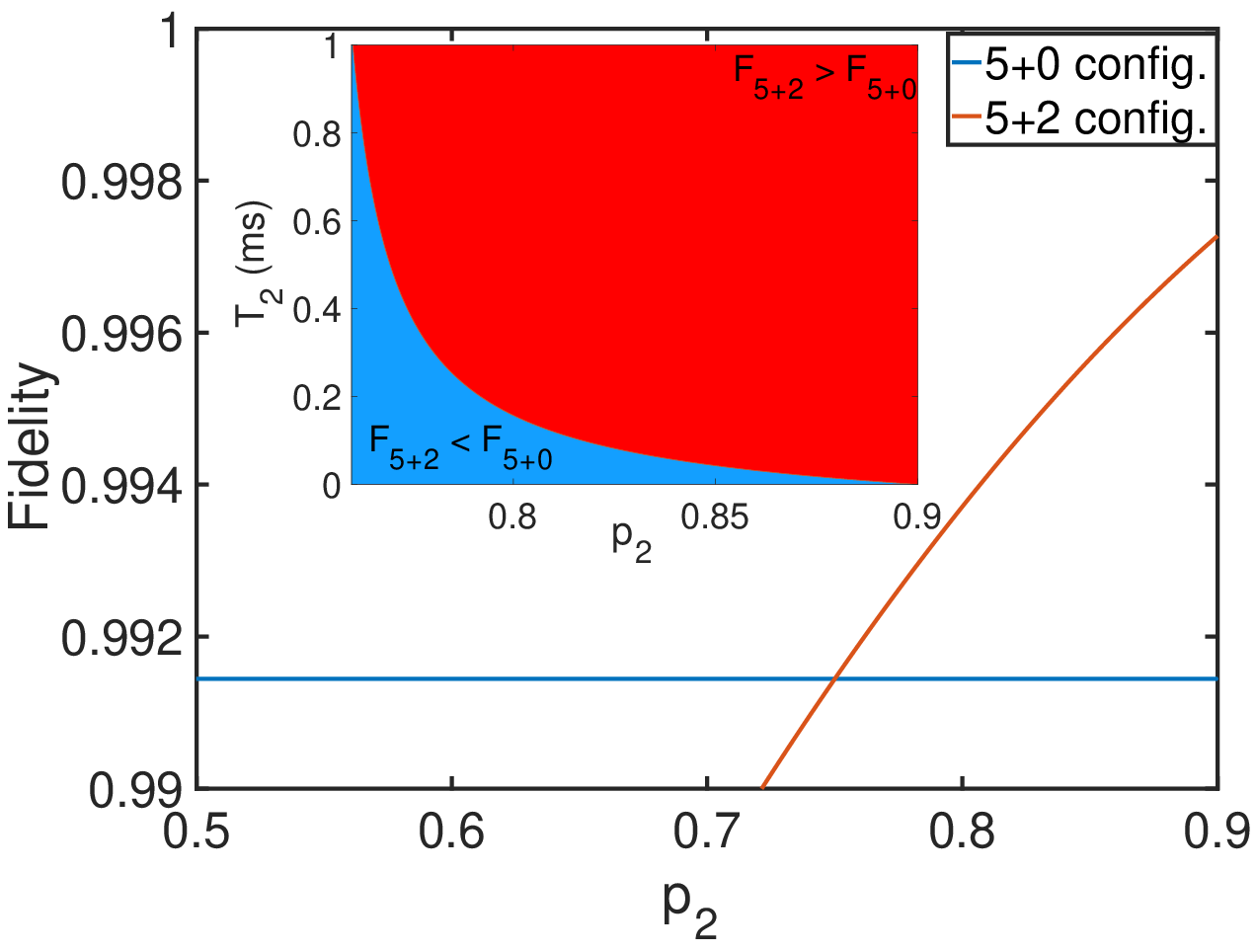}
\par\end{centering}
\caption{\label{fig:Fidelity-greedy}Fidelity of the transmitted state corresponding
to the 5+2 configuration (red curve) and 5+0 configuration (blue line).
The inset shows the optimal configurations when the fidelities are
affected by the coherence time, $T{{}_2}$, by plotting $T{{}_2}$
versus $p{{}_2}$.}
\end{figure}

Next we consider the case in which the router selects the best configuration
in the greedy regime. For instance, if user 1 requires the highest
fidelity, the router will assign user 1 most of the resources traveling
along path 1. To this end we then compare the fidelity of \textcolor{black}{the
configuration 5+2}\textcolor{red}{{} }with the fidelity of the configuration
5+0 as shown in Fig. \ref{fig:Fidelity-greedy}
. Interestingly, the advantage of using higher dimensional code is
achieved for $p_{2}>0.75.$ For lower values of $p_{2}$ it is more
advantageous to encode the state in a lower dimensional code. This
might seem a bit counterintuitive because the 7-dimensional QRS code
can tolerate the loss of three qudits while the 5-dimensional code
can tolerate the loss of only 2 qudits. Therefore, one might think
that at $p_{2}=0$, i.e., when the two qudits traveling in path 2
are lost, the fidelity of the encoded state using the 7-dimensional
code is equal to the fidelity of the 5 dimensional code. This not
the case, because when the two qudits in path 2 are lost the 7-dimensional
code can tolerate the loss of one qudit traveling in path 1 (or, obviously,
zero qudit). Therefore the contribution to the fidelity in this case
are proportional to $p_{1}^{5}$ (no loss in path 1) and $p_{1}^{4}(1-p_{1})$
(1 qudit loss in path 1). On the other hand, when the state has been
encoded with a 5-dimensional QRS code with all qudits traveling in
path 1, the fidelity will also have a contribution proportional to
$p_{1}^{3}(1-p_{1})^{2}$ in addition to the above contributions.
These results are obtained for ideal QMs at the receiver's node, in
which the coherence time is assumed to be infinite. 

\subsection{The effect of finite coherence of QMs}

Let us now consider the case in which the QMs at the receiver's node
are not perfect and have a finite coherence time $T_{2}.$ This can
drastically change the optimal assignment made by the router as illustrated
in the following example. Let us assume we are in the greedy regime,
in which a single user demands \textcolor{black}{that its transmitted
state achieves the highest possible fidelity} with the available resources.
Looking at Fig. \ref{fig:Fidelity-greedy} one notices that at $p_{2}<0.72$
the 5+0 configuration is the most advantageous. We want to determine
then for which values of $T_{2}$ the 5+2 configuration has a better
performance than the 5+0 configuration for $p_{2}>0.75.$ The inset
of fig. \ref{fig:Fidelity-greedy} shows the red(blue) area delimited
by the coherence time $T_{2}$ and by $p_{2}$ to obtain a fidelity
of the transmitted state higher(lower) in the 5+2 configuration compared
to the 5+0 configuration. For $T_{2}>1$ ms the asymptotic regime
is achieved and the coherence time only slightly affects the fidelity
in the 5+2 configuration. However, for $T_{2}<1$ ms one can see that
higher values of $p_{2}$ are needed to guarantee that the 5+2 configuration
gives an advantage compared to the 5+0 configuration. 

Similarly the inset of Fig. \ref{fig:Difference-of-thefidelities}
shows that the coherence time plays a relevant role in assigning the
channels in the restricted regime as well. First one can notice that
at $T_{2}=1$ ms the asymptotic regime is achieved as well as in the
greedy regime. Next, at lower values of $T_{2}>0.2$ ms, the 3+4 and
2+1 configurations (red area) have more similar fidelities than the
2+5 and 3+0 configuration (black area). It is\textcolor{black}{{} noteworthy
to notic}e that at $T_{2}<0.2$ ms the 3+2 and 2+3 have much more
similar fidelities than all the other configurations under consideration. 

\textcolor{black}{These results highlight that coherence time can
play a critical role in determining the optimal assignment of channels.
Therefore, it is essential that the router be informed of the coherence
time of the QMs used at the remote end to ensure it meets the users\textquoteright{}
performance expectations.}

\section{Conclusion}

\textcolor{black}{In this work, we have introduced the concept of
Quality of Service (QoS) tailored for quantum and aggregated quantum
networks, providing a qualitative description of their requirements
across several quantum communication applications. We proposed a routing
protocol for the optimal assignment of quantum channels connecting
two remote nodes with multiple users, illustrating its effectiveness
through two representative scenarios.}

\textcolor{black}{Our analysis showed that when users transmit unencoded
quantum states, the coherence time of quantum memories does not impact
optimal channel assignment within the considered regimes, though a
threshold coherence time exists below which quantum aggregation is
no longer advantageous. In contrast, when employing quantum error
correction---specifically, the quantum Reed-Solomon code---the optimal
assignments strongly depend on both channel transmission probabilities
and the coherence times of quantum memories at the receiver node.
We quantified coherence time thresholds above which decoherence effects
become negligible, demonstrating that current quantum memory technologies,
such as ion qubits, superconducting cavities, and ensemble-based systems,
are suitable for such networks.}

\textcolor{black}{Although our model focuses on quantum networks without
entanglement distribution, it draws inspiration from classical wireless
mesh networks \citep{MeshN1,MeshN2}, leveraging concepts like decentralized
routing, multi-user coordination, and QoS-aware path selection to
guide protocol design. We identified bandwidth and loss as crucial
QoS parameters for optimizing channel assignment in small-scale aggregated
networks, and suggest that incorporating additional metrics such as
delay and jitter could further improve performance.}

\textcolor{black}{Overall, this work establishes QoS as a vital figure
of merit for quantum networks and provides a foundation for optimizing
their design and operation in future implementations.}
\begin{acknowledgments}
We thank Mohsen Razavi for the interesting discussion.\textcolor{black}{{}
This project was made possible through the support of the Moonshot
R\&D Program Grants JPMJMS2061 \& JPMJMS226C, JSPS KAKENHI Grant Nos.
21H04880 and 24K07485.}
\end{acknowledgments}

\bibliography{bib1}

\begin{thebibliography}{47}%
\makeatletter
\providecommand \@ifxundefined [1]{%
 \@ifx{#1\undefined}
}%
\providecommand \@ifnum [1]{%
 \ifnum #1\expandafter \@firstoftwo
 \else \expandafter \@secondoftwo
 \fi
}%
\providecommand \@ifx [1]{%
 \ifx #1\expandafter \@firstoftwo
 \else \expandafter \@secondoftwo
 \fi
}%
\providecommand \natexlab [1]{#1}%
\providecommand \enquote  [1]{``#1''}%
\providecommand \bibnamefont  [1]{#1}%
\providecommand \bibfnamefont [1]{#1}%
\providecommand \citenamefont [1]{#1}%
\providecommand \href@noop [0]{\@secondoftwo}%
\providecommand \href [0]{\begingroup \@sanitize@url \@href}%
\providecommand \@href[1]{\@@startlink{#1}\@@href}%
\providecommand \@@href[1]{\endgroup#1\@@endlink}%
\providecommand \@sanitize@url [0]{\catcode `\\12\catcode `\$12\catcode
  `\&12\catcode `\#12\catcode `\^12\catcode `\_12\catcode `\%12\relax}%
\providecommand \@@startlink[1]{}%
\providecommand \@@endlink[0]{}%
\providecommand \url  [0]{\begingroup\@sanitize@url \@url }%
\providecommand \@url [1]{\endgroup\@href {#1}{\urlprefix }}%
\providecommand \urlprefix  [0]{URL }%
\providecommand \Eprint [0]{\href }%
\providecommand \doibase [0]{https://doi.org/}%
\providecommand \selectlanguage [0]{\@gobble}%
\providecommand \bibinfo  [0]{\@secondoftwo}%
\providecommand \bibfield  [0]{\@secondoftwo}%
\providecommand \translation [1]{[#1]}%
\providecommand \BibitemOpen [0]{}%
\providecommand \bibitemStop [0]{}%
\providecommand \bibitemNoStop [0]{.\EOS\space}%
\providecommand \EOS [0]{\spacefactor3000\relax}%
\providecommand \BibitemShut  [1]{\csname bibitem#1\endcsname}%
\let\auto@bib@innerbib\@empty
\bibitem [{\citenamefont {Kozlowski}\ and\ \citenamefont
  {Wehner}(2019)}]{QNET1}%
  \BibitemOpen
  \bibfield  {author} {\bibinfo {author} {\bibfnamefont {W.}~\bibnamefont
  {Kozlowski}}\ and\ \bibinfo {author} {\bibfnamefont {S.}~\bibnamefont
  {Wehner}},\ }\bibfield  {title} {\bibinfo {title} {Towards large-scale
  quantum networks},\ }\href@noop {} {\bibfield  {journal} {\bibinfo  {journal}
  {In Proceedings of the 6th ACM International Conference on Nanoscale
  Computing and Communication}\ } (\bibinfo {year} {2019})}\BibitemShut
  {NoStop}%
\bibitem [{\citenamefont {Childress}\ and\ \citenamefont
  {Hanson}(2013)}]{NVnetworks1}%
  \BibitemOpen
  \bibfield  {author} {\bibinfo {author} {\bibfnamefont {L.}~\bibnamefont
  {Childress}}\ and\ \bibinfo {author} {\bibfnamefont {R.}~\bibnamefont
  {Hanson}},\ }\bibfield  {title} {\bibinfo {title} {Diamond nv centers for
  quantum computing and quantum networks},\ }\href@noop {} {\bibfield
  {journal} {\bibinfo  {journal} {MRS Bulletin}\ }\textbf {\bibinfo {volume}
  {38}},\ \bibinfo {pages} {134} (\bibinfo {year} {2013})}\BibitemShut
  {NoStop}%
\bibitem [{\citenamefont {Blok}\ \emph {et~al.}(2015)\citenamefont {Blok},
  \citenamefont {Kalb}, \citenamefont {Reiserer}, \citenamefont {Taminiau},\
  and\ \citenamefont {Hanson}}]{NVnetworks2}%
  \BibitemOpen
  \bibfield  {author} {\bibinfo {author} {\bibfnamefont {M.~S.}\ \bibnamefont
  {Blok}}, \bibinfo {author} {\bibfnamefont {N.}~\bibnamefont {Kalb}}, \bibinfo
  {author} {\bibfnamefont {A.}~\bibnamefont {Reiserer}}, \bibinfo {author}
  {\bibfnamefont {T.~H.}\ \bibnamefont {Taminiau}},\ and\ \bibinfo {author}
  {\bibfnamefont {R.}~\bibnamefont {Hanson}},\ }\bibfield  {title} {\bibinfo
  {title} {Towards quantum networks of single spins: analysis of a quantum
  memory with an optical interface in diamond},\ }\href@noop {} {\bibfield
  {journal} {\bibinfo  {journal} {Faraday Discuss.}\ }\textbf {\bibinfo
  {volume} {184}},\ \bibinfo {pages} {173} (\bibinfo {year}
  {2015})}\BibitemShut {NoStop}%
\bibitem [{\citenamefont {Kimble}(2008)}]{Q_internet}%
  \BibitemOpen
  \bibfield  {author} {\bibinfo {author} {\bibfnamefont {H.~J.}\ \bibnamefont
  {Kimble}},\ }\bibfield  {title} {\bibinfo {title} {The quantum internet},\
  }\href@noop {} {\bibfield  {journal} {\bibinfo  {journal} {Nature}\ }\textbf
  {\bibinfo {volume} {453}},\ \bibinfo {pages} {1023} (\bibinfo {year}
  {2008})}\BibitemShut {NoStop}%
\bibitem [{\citenamefont {Arute}\ \emph {et~al.}(2019)\citenamefont {Arute},
  \citenamefont {Arya}, \citenamefont {Babbush}, \citenamefont {Bacon},
  \citenamefont {Bardin}, \citenamefont {Barends}, \citenamefont {Biswas},
  \citenamefont {Boixo}, \citenamefont {Brandao}, \citenamefont {Buell},
  \citenamefont {Burkett}, \citenamefont {Chen}, \citenamefont {Chen},
  \citenamefont {Chiaro}, \citenamefont {Collins}, \citenamefont {Courtney},
  \citenamefont {Dunsworth}, \citenamefont {Farhi}, \citenamefont {Foxen},
  \citenamefont {Fowler}, \citenamefont {Gidney}, \citenamefont {Giustina},
  \citenamefont {Graff}, \citenamefont {Guerin}, \citenamefont {Habegger},
  \citenamefont {Harrigan}, \citenamefont {Hartmann}, \citenamefont {Ho},
  \citenamefont {Hoffmann}, \citenamefont {Huang}, \citenamefont {Humble},
  \citenamefont {I.}, \citenamefont {Jeffrey}, \citenamefont {Jiang},
  \citenamefont {Kafri}, \citenamefont {Kechedzhi}, \citenamefont {Kelly},
  \citenamefont {Klimov}, \citenamefont {Knysh}, \citenamefont {Korotkov},
  \citenamefont {Kostritsa}, \citenamefont {Landhuis}, \citenamefont
  {Lindmark}, \citenamefont {Lucero}, \citenamefont {Lyakh}, \citenamefont
  {Mandrà}, \citenamefont {McClean}, \citenamefont {McEwen}, \citenamefont
  {Megrant}, \citenamefont {Mi}, \citenamefont {Michielsen}, \citenamefont
  {Mohseni}, \citenamefont {Mutus}, \citenamefont {Naaman}, \citenamefont
  {Neeley}, \citenamefont {Neill}, \citenamefont {Niu}, \citenamefont {Ostby},
  \citenamefont {Petukhov}, \citenamefont {Platt}, \citenamefont {Quintana},
  \citenamefont {Rieffel}, \citenamefont {Roushan}, \citenamefont {Rubin},
  \citenamefont {Sank}, \citenamefont {Satzinger}, \citenamefont {Smelyanskiy},
  \citenamefont {Sung}, \citenamefont {Trevithick}, \citenamefont
  {Vainsencher}, \citenamefont {Villalonga}, \citenamefont {White},
  \citenamefont {Yao}, \citenamefont {Yeh}, \citenamefont {Zalcman},
  \citenamefont {Neven},\ and\ \citenamefont {Martinis}}]{QC_google}%
  \BibitemOpen
  \bibfield  {author} {\bibinfo {author} {\bibfnamefont {F.}~\bibnamefont
  {Arute}}, \bibinfo {author} {\bibfnamefont {K.}~\bibnamefont {Arya}},
  \bibinfo {author} {\bibfnamefont {R.}~\bibnamefont {Babbush}}, \bibinfo
  {author} {\bibfnamefont {D.}~\bibnamefont {Bacon}}, \bibinfo {author}
  {\bibfnamefont {J.~C.}\ \bibnamefont {Bardin}}, \bibinfo {author}
  {\bibfnamefont {R.}~\bibnamefont {Barends}}, \bibinfo {author} {\bibfnamefont
  {R.}~\bibnamefont {Biswas}}, \bibinfo {author} {\bibfnamefont
  {S.}~\bibnamefont {Boixo}}, \bibinfo {author} {\bibfnamefont {F.~G. S.~L.}\
  \bibnamefont {Brandao}}, \bibinfo {author} {\bibfnamefont {D.~A.}\
  \bibnamefont {Buell}}, \bibinfo {author} {\bibfnamefont {B.}~\bibnamefont
  {Burkett}}, \bibinfo {author} {\bibfnamefont {Y.}~\bibnamefont {Chen}},
  \bibinfo {author} {\bibfnamefont {Z.}~\bibnamefont {Chen}}, \bibinfo {author}
  {\bibfnamefont {B.}~\bibnamefont {Chiaro}}, \bibinfo {author} {\bibfnamefont
  {R.}~\bibnamefont {Collins}}, \bibinfo {author} {\bibfnamefont
  {W.}~\bibnamefont {Courtney}}, \bibinfo {author} {\bibfnamefont
  {A.}~\bibnamefont {Dunsworth}}, \bibinfo {author} {\bibfnamefont
  {E.}~\bibnamefont {Farhi}}, \bibinfo {author} {\bibfnamefont
  {B.}~\bibnamefont {Foxen}}, \bibinfo {author} {\bibfnamefont
  {A.}~\bibnamefont {Fowler}}, \bibinfo {author} {\bibfnamefont
  {C.}~\bibnamefont {Gidney}}, \bibinfo {author} {\bibfnamefont
  {M.}~\bibnamefont {Giustina}}, \bibinfo {author} {\bibfnamefont
  {R.}~\bibnamefont {Graff}}, \bibinfo {author} {\bibfnamefont
  {K.}~\bibnamefont {Guerin}}, \bibinfo {author} {\bibfnamefont
  {S.}~\bibnamefont {Habegger}}, \bibinfo {author} {\bibfnamefont {M.~P.}\
  \bibnamefont {Harrigan}}, \bibinfo {author} {\bibfnamefont {M.~J.}\
  \bibnamefont {Hartmann}}, \bibinfo {author} {\bibfnamefont {A.}~\bibnamefont
  {Ho}}, \bibinfo {author} {\bibfnamefont {M.}~\bibnamefont {Hoffmann}},
  \bibinfo {author} {\bibfnamefont {T.}~\bibnamefont {Huang}}, \bibinfo
  {author} {\bibfnamefont {T.~S.}\ \bibnamefont {Humble}}, \bibinfo {author}
  {\bibfnamefont {S.~V.}\ \bibnamefont {I.}}, \bibinfo {author} {\bibfnamefont
  {E.}~\bibnamefont {Jeffrey}}, \bibinfo {author} {\bibfnamefont
  {Z.}~\bibnamefont {Jiang}}, \bibinfo {author} {\bibfnamefont
  {D.}~\bibnamefont {Kafri}}, \bibinfo {author} {\bibfnamefont
  {K.}~\bibnamefont {Kechedzhi}}, \bibinfo {author} {\bibfnamefont
  {J.}~\bibnamefont {Kelly}}, \bibinfo {author} {\bibfnamefont {P.~V.}\
  \bibnamefont {Klimov}}, \bibinfo {author} {\bibfnamefont {S.}~\bibnamefont
  {Knysh}}, \bibinfo {author} {\bibfnamefont {A.}~\bibnamefont {Korotkov}},
  \bibinfo {author} {\bibfnamefont {F.}~\bibnamefont {Kostritsa}}, \bibinfo
  {author} {\bibfnamefont {D.}~\bibnamefont {Landhuis}}, \bibinfo {author}
  {\bibfnamefont {M.}~\bibnamefont {Lindmark}}, \bibinfo {author}
  {\bibfnamefont {E.}~\bibnamefont {Lucero}}, \bibinfo {author} {\bibfnamefont
  {D.}~\bibnamefont {Lyakh}}, \bibinfo {author} {\bibfnamefont
  {S.}~\bibnamefont {Mandrà}}, \bibinfo {author} {\bibfnamefont {J.~R.}\
  \bibnamefont {McClean}}, \bibinfo {author} {\bibfnamefont {M.}~\bibnamefont
  {McEwen}}, \bibinfo {author} {\bibfnamefont {A.}~\bibnamefont {Megrant}},
  \bibinfo {author} {\bibfnamefont {X.}~\bibnamefont {Mi}}, \bibinfo {author}
  {\bibfnamefont {K.}~\bibnamefont {Michielsen}}, \bibinfo {author}
  {\bibfnamefont {M.}~\bibnamefont {Mohseni}}, \bibinfo {author} {\bibfnamefont
  {J.}~\bibnamefont {Mutus}}, \bibinfo {author} {\bibfnamefont
  {O.}~\bibnamefont {Naaman}}, \bibinfo {author} {\bibfnamefont
  {M.}~\bibnamefont {Neeley}}, \bibinfo {author} {\bibfnamefont
  {C.}~\bibnamefont {Neill}}, \bibinfo {author} {\bibfnamefont {M.~Y.}\
  \bibnamefont {Niu}}, \bibinfo {author} {\bibfnamefont {E.}~\bibnamefont
  {Ostby}}, \bibinfo {author} {\bibfnamefont {A.}~\bibnamefont {Petukhov}},
  \bibinfo {author} {\bibfnamefont {J.~C.}\ \bibnamefont {Platt}}, \bibinfo
  {author} {\bibfnamefont {C.}~\bibnamefont {Quintana}}, \bibinfo {author}
  {\bibfnamefont {E.~G.}\ \bibnamefont {Rieffel}}, \bibinfo {author}
  {\bibfnamefont {P.}~\bibnamefont {Roushan}}, \bibinfo {author} {\bibfnamefont
  {N.~C.}\ \bibnamefont {Rubin}}, \bibinfo {author} {\bibfnamefont
  {D.}~\bibnamefont {Sank}}, \bibinfo {author} {\bibfnamefont {K.~J.}\
  \bibnamefont {Satzinger}}, \bibinfo {author} {\bibfnamefont {V.}~\bibnamefont
  {Smelyanskiy}}, \bibinfo {author} {\bibfnamefont {K.~J.}\ \bibnamefont
  {Sung}}, \bibinfo {author} {\bibfnamefont {M.~D.}\ \bibnamefont
  {Trevithick}}, \bibinfo {author} {\bibfnamefont {A.}~\bibnamefont
  {Vainsencher}}, \bibinfo {author} {\bibfnamefont {B.}~\bibnamefont
  {Villalonga}}, \bibinfo {author} {\bibfnamefont {T.}~\bibnamefont {White}},
  \bibinfo {author} {\bibfnamefont {Z.~J.}\ \bibnamefont {Yao}}, \bibinfo
  {author} {\bibfnamefont {P.}~\bibnamefont {Yeh}}, \bibinfo {author}
  {\bibfnamefont {A.}~\bibnamefont {Zalcman}}, \bibinfo {author} {\bibfnamefont
  {H.}~\bibnamefont {Neven}},\ and\ \bibinfo {author} {\bibfnamefont {J.~M.}\
  \bibnamefont {Martinis}},\ }\bibfield  {title} {\bibinfo {title} {Quantum
  supremacy using a programmable superconducting processor},\ }\href@noop {}
  {\bibfield  {journal} {\bibinfo  {journal} {Nature}\ }\textbf {\bibinfo
  {volume} {574}},\ \bibinfo {pages} {505} (\bibinfo {year}
  {2019})}\BibitemShut {NoStop}%
\bibitem [{\citenamefont {Munro}\ \emph {et~al.}(2015)\citenamefont {Munro},
  \citenamefont {Azuma}, \citenamefont {Tamaki},\ and\ \citenamefont
  {Nemoto}}]{QC_Bill}%
  \BibitemOpen
  \bibfield  {author} {\bibinfo {author} {\bibfnamefont {W.~J.}\ \bibnamefont
  {Munro}}, \bibinfo {author} {\bibfnamefont {K.}~\bibnamefont {Azuma}},
  \bibinfo {author} {\bibfnamefont {K.}~\bibnamefont {Tamaki}},\ and\ \bibinfo
  {author} {\bibfnamefont {K.}~\bibnamefont {Nemoto}},\ }\bibfield  {title}
  {\bibinfo {title} {Inside quantum repeaters},\ }\href@noop {} {\bibfield
  {journal} {\bibinfo  {journal} {IEEE Journal of Selected Topics in Quantum
  Electronics}\ }\textbf {\bibinfo {volume} {21}},\ \bibinfo {pages} {6400813}
  (\bibinfo {year} {2015})}\BibitemShut {NoStop}%
\bibitem [{\citenamefont {Nielsen}\ and\ \citenamefont
  {Chuang}(2000)}]{Qcomp1}%
  \BibitemOpen
  \bibfield  {author} {\bibinfo {author} {\bibfnamefont {M.}~\bibnamefont
  {Nielsen}}\ and\ \bibinfo {author} {\bibfnamefont {I.}~\bibnamefont
  {Chuang}},\ }\href@noop {} {\emph {\bibinfo {title} {Quantum Computation and
  Quantum Information}}}\ (\bibinfo  {publisher} {Cambridge University Press,
  Cambridge},\ \bibinfo {year} {2000})\BibitemShut {NoStop}%
\bibitem [{\citenamefont {Bennett}\ and\ \citenamefont
  {DiVincenzo}(2000)}]{Qcomp2}%
  \BibitemOpen
  \bibfield  {author} {\bibinfo {author} {\bibfnamefont {C.}~\bibnamefont
  {Bennett}}\ and\ \bibinfo {author} {\bibfnamefont {D.}~\bibnamefont
  {DiVincenzo}},\ }\bibfield  {title} {\bibinfo {title} {Quantum information
  and computation},\ }\href@noop {} {\bibfield  {journal} {\bibinfo  {journal}
  {Nature}\ }\textbf {\bibinfo {volume} {404}},\ \bibinfo {pages} {247}
  (\bibinfo {year} {2000})}\BibitemShut {NoStop}%
\bibitem [{\citenamefont {Raussendorf}\ and\ \citenamefont
  {Briegel}(2001)}]{Quantum_comp2}%
  \BibitemOpen
  \bibfield  {author} {\bibinfo {author} {\bibfnamefont {R.}~\bibnamefont
  {Raussendorf}}\ and\ \bibinfo {author} {\bibfnamefont {H.~J.}\ \bibnamefont
  {Briegel}},\ }\bibfield  {title} {\bibinfo {title} {A one-way quantum
  computer},\ }\href@noop {} {\bibfield  {journal} {\bibinfo  {journal} {Phys.
  Rev. Lett.}\ }\textbf {\bibinfo {volume} {86}},\ \bibinfo {pages} {5188}
  (\bibinfo {year} {2001})}\BibitemShut {NoStop}%
\bibitem [{\citenamefont {Knill}(2005)}]{Quantum_comp3}%
  \BibitemOpen
  \bibfield  {author} {\bibinfo {author} {\bibfnamefont {E.}~\bibnamefont
  {Knill}},\ }\bibfield  {title} {\bibinfo {title} {Quantum computing with
  realistically noisy devices},\ }\href@noop {} {\bibfield  {journal} {\bibinfo
   {journal} {Nature}\ }\textbf {\bibinfo {volume} {434}},\ \bibinfo {pages}
  {39} (\bibinfo {year} {2005})}\BibitemShut {NoStop}%
\bibitem [{\citenamefont {Duan}\ and\ \citenamefont
  {Raussendorf}(2005)}]{Quantum_comp4}%
  \BibitemOpen
  \bibfield  {author} {\bibinfo {author} {\bibfnamefont {L.-M.}\ \bibnamefont
  {Duan}}\ and\ \bibinfo {author} {\bibfnamefont {R.}~\bibnamefont
  {Raussendorf}},\ }\bibfield  {title} {\bibinfo {title} {Scalable photonic
  quantum computation through cavity-assisted interactions},\ }\href@noop {}
  {\bibfield  {journal} {\bibinfo  {journal} {Phys. Rev. Lett.}\ }\textbf
  {\bibinfo {volume} {95}},\ \bibinfo {pages} {080503} (\bibinfo {year}
  {2005})}\BibitemShut {NoStop}%
\bibitem [{\citenamefont {Hwang}(2003)}]{QKD03}%
  \BibitemOpen
  \bibfield  {author} {\bibinfo {author} {\bibfnamefont {W.-Y.}\ \bibnamefont
  {Hwang}},\ }\bibfield  {title} {\bibinfo {title} {Quantum key distribution
  with high loss: Toward global secure communication},\ }\href@noop {}
  {\bibfield  {journal} {\bibinfo  {journal} {Phys. Rev. Lett.}\ }\textbf
  {\bibinfo {volume} {91}},\ \bibinfo {pages} {057901} (\bibinfo {year}
  {2003})}\BibitemShut {NoStop}%
\bibitem [{\citenamefont {Ekert}(1991)}]{QKD1}%
  \BibitemOpen
  \bibfield  {author} {\bibinfo {author} {\bibfnamefont {A.~K.}\ \bibnamefont
  {Ekert}},\ }\bibfield  {title} {\bibinfo {title} {Quantum cryptography based
  on bell's theorem},\ }\href@noop {} {\bibfield  {journal} {\bibinfo
  {journal} {Phys. Rev. Lett.}\ }\textbf {\bibinfo {volume} {67}},\ \bibinfo
  {pages} {661} (\bibinfo {year} {1991})}\BibitemShut {NoStop}%
\bibitem [{\citenamefont {Qiu}(2014)}]{QKD2000}%
  \BibitemOpen
  \bibfield  {author} {\bibinfo {author} {\bibfnamefont {J.}~\bibnamefont
  {Qiu}},\ }\bibfield  {title} {\bibinfo {title} {Quantum communications leap
  out of the lab},\ }\href@noop {} {\bibfield  {journal} {\bibinfo  {journal}
  {Nature}\ }\textbf {\bibinfo {volume} {508}},\ \bibinfo {pages} {441}
  (\bibinfo {year} {2014})}\BibitemShut {NoStop}%
\bibitem [{\citenamefont {Bunandar}\ \emph
  {et~al.}(2018{\natexlab{a}})\citenamefont {Bunandar}, \citenamefont
  {Lentine}, \citenamefont {Lee}, \citenamefont {Cai}, \citenamefont {Long},
  \citenamefont {Boynton}, \citenamefont {Martinez}, \citenamefont {DeRose},
  \citenamefont {Chen}, \citenamefont {Grein}, \citenamefont {Trotter},
  \citenamefont {Starbuck}, \citenamefont {Pomerene}, \citenamefont {Hamilton},
  \citenamefont {Wong}, \citenamefont {Camacho}, \citenamefont {Davids},
  \citenamefont {Urayama},\ and\ \citenamefont {Englund}}]{QKD_transmission}%
  \BibitemOpen
  \bibfield  {author} {\bibinfo {author} {\bibfnamefont {D.}~\bibnamefont
  {Bunandar}}, \bibinfo {author} {\bibfnamefont {A.}~\bibnamefont {Lentine}},
  \bibinfo {author} {\bibfnamefont {C.}~\bibnamefont {Lee}}, \bibinfo {author}
  {\bibfnamefont {H.}~\bibnamefont {Cai}}, \bibinfo {author} {\bibfnamefont
  {C.~M.}\ \bibnamefont {Long}}, \bibinfo {author} {\bibfnamefont
  {N.}~\bibnamefont {Boynton}}, \bibinfo {author} {\bibfnamefont
  {N.}~\bibnamefont {Martinez}}, \bibinfo {author} {\bibfnamefont
  {C.}~\bibnamefont {DeRose}}, \bibinfo {author} {\bibfnamefont
  {C.}~\bibnamefont {Chen}}, \bibinfo {author} {\bibfnamefont {M.}~\bibnamefont
  {Grein}}, \bibinfo {author} {\bibfnamefont {D.}~\bibnamefont {Trotter}},
  \bibinfo {author} {\bibfnamefont {A.}~\bibnamefont {Starbuck}}, \bibinfo
  {author} {\bibfnamefont {A.}~\bibnamefont {Pomerene}}, \bibinfo {author}
  {\bibfnamefont {S.}~\bibnamefont {Hamilton}}, \bibinfo {author}
  {\bibfnamefont {F.~N.~C.}\ \bibnamefont {Wong}}, \bibinfo {author}
  {\bibfnamefont {R.}~\bibnamefont {Camacho}}, \bibinfo {author} {\bibfnamefont
  {P.}~\bibnamefont {Davids}}, \bibinfo {author} {\bibfnamefont
  {J.}~\bibnamefont {Urayama}},\ and\ \bibinfo {author} {\bibfnamefont
  {D.}~\bibnamefont {Englund}},\ }\bibfield  {title} {\bibinfo {title}
  {Metropolitan quantum key distribution with silicon photonics},\ }\href@noop
  {} {\bibfield  {journal} {\bibinfo  {journal} {Phys. Rev. X}\ }\textbf
  {\bibinfo {volume} {8}},\ \bibinfo {pages} {12} (\bibinfo {year}
  {2018}{\natexlab{a}})}\BibitemShut {NoStop}%
\bibitem [{\citenamefont {Simon}\ \emph {et~al.}(2014)\citenamefont {Simon},
  \citenamefont {Jaeger},\ and\ \citenamefont {Sergienko}}]{QImaging2}%
  \BibitemOpen
  \bibfield  {author} {\bibinfo {author} {\bibfnamefont {D.~S.}\ \bibnamefont
  {Simon}}, \bibinfo {author} {\bibfnamefont {G.}~\bibnamefont {Jaeger}},\ and\
  \bibinfo {author} {\bibfnamefont {A.~V.}\ \bibnamefont {Sergienko}},\
  }\bibfield  {title} {\bibinfo {title} {Quantum metrology, imaging, and
  communication book},\ }\href@noop {} {\bibfield  {journal} {\bibinfo
  {journal} {Int. J. Quantum Inform.}\ }\textbf {\bibinfo {volume} {12}},\
  \bibinfo {pages} {1430004} (\bibinfo {year} {2014})}\BibitemShut {NoStop}%
\bibitem [{\citenamefont {Lugiato}\ \emph {et~al.}(2002)\citenamefont
  {Lugiato}, \citenamefont {Gatti},\ and\ \citenamefont
  {Brambilla}}]{QImaging3}%
  \BibitemOpen
  \bibfield  {author} {\bibinfo {author} {\bibfnamefont {L.~A.}\ \bibnamefont
  {Lugiato}}, \bibinfo {author} {\bibfnamefont {A.}~\bibnamefont {Gatti}},\
  and\ \bibinfo {author} {\bibfnamefont {E.}~\bibnamefont {Brambilla}},\
  }\bibfield  {title} {\bibinfo {title} {Quantum imaging},\ }\href@noop {}
  {\bibfield  {journal} {\bibinfo  {journal} {J. Opt. B}\ }\textbf {\bibinfo
  {volume} {4}},\ \bibinfo {pages} {176} (\bibinfo {year} {2002})}\BibitemShut
  {NoStop}%
\bibitem [{\citenamefont {Dogen}\ \emph {et~al.}(2017)\citenamefont {Dogen},
  \citenamefont {Reinhard},\ and\ \citenamefont {Cappellaro}}]{QSensing}%
  \BibitemOpen
  \bibfield  {author} {\bibinfo {author} {\bibfnamefont {C.~L.}\ \bibnamefont
  {Dogen}}, \bibinfo {author} {\bibfnamefont {F.}~\bibnamefont {Reinhard}},\
  and\ \bibinfo {author} {\bibfnamefont {P.}~\bibnamefont {Cappellaro}},\
  }\bibfield  {title} {\bibinfo {title} {Quantum sensing},\ }\href@noop {}
  {\bibfield  {journal} {\bibinfo  {journal} {Rev. Mod. Phys.}\ }\textbf
  {\bibinfo {volume} {89}},\ \bibinfo {pages} {035002} (\bibinfo {year}
  {2017})}\BibitemShut {NoStop}%
\bibitem [{\citenamefont {Fanizza}\ \emph {et~al.}(2020)\citenamefont
  {Fanizza}, \citenamefont {Kianvash},\ and\ \citenamefont
  {Giovannetti}}]{q_capacity1}%
  \BibitemOpen
  \bibfield  {author} {\bibinfo {author} {\bibfnamefont {M.}~\bibnamefont
  {Fanizza}}, \bibinfo {author} {\bibfnamefont {F.}~\bibnamefont {Kianvash}},\
  and\ \bibinfo {author} {\bibfnamefont {V.}~\bibnamefont {Giovannetti}},\
  }\bibfield  {title} {\bibinfo {title} {Quantum flags and new bounds on the
  quantum capacity of the depolarizing channel},\ }\href@noop {} {\bibfield
  {journal} {\bibinfo  {journal} {Phys. Rev. Lett.}\ }\textbf {\bibinfo
  {volume} {125}},\ \bibinfo {pages} {020503} (\bibinfo {year}
  {2020})}\BibitemShut {NoStop}%
\bibitem [{\citenamefont {Rosati}\ \emph {et~al.}(2018)\citenamefont {Rosati},
  \citenamefont {Mari},\ and\ \citenamefont {Giovannetti}}]{q_capacity2}%
  \BibitemOpen
  \bibfield  {author} {\bibinfo {author} {\bibfnamefont {M.}~\bibnamefont
  {Rosati}}, \bibinfo {author} {\bibfnamefont {A.}~\bibnamefont {Mari}},\ and\
  \bibinfo {author} {\bibfnamefont {V.}~\bibnamefont {Giovannetti}},\
  }\bibfield  {title} {\bibinfo {title} {Narrow bounds for the quantum capacity
  of thermal attenuators},\ }\href@noop {} {\bibfield  {journal} {\bibinfo
  {journal} {Nat. Commun.}\ }\textbf {\bibinfo {volume} {9}},\ \bibinfo {pages}
  {4339} (\bibinfo {year} {2018})}\BibitemShut {NoStop}%
\bibitem [{\citenamefont {Shirokov}(2017)}]{q_capacity3}%
  \BibitemOpen
  \bibfield  {author} {\bibinfo {author} {\bibfnamefont {M.~E.}\ \bibnamefont
  {Shirokov}},\ }\bibfield  {title} {\bibinfo {title} {Uniform continuity
  bounds for characteristics of multipartite quantum systems},\ }\href@noop {}
  {\bibfield  {journal} {\bibinfo  {journal} {Journal of Mathematical Physics}\
  }\textbf {\bibinfo {volume} {58}},\ \bibinfo {pages} {102202} (\bibinfo
  {year} {2017})}\BibitemShut {NoStop}%
\bibitem [{\citenamefont {Skrzypczyk}\ and\ \citenamefont
  {Wehner}(2021)}]{Wehener_QoS}%
  \BibitemOpen
  \bibfield  {author} {\bibinfo {author} {\bibfnamefont {M.}~\bibnamefont
  {Skrzypczyk}}\ and\ \bibinfo {author} {\bibfnamefont {S.}~\bibnamefont
  {Wehner}},\ }\bibfield  {title} {\bibinfo {title} {An architecture for
  meeting quality-of-service requirements in multi-user quantum networks},\
  }\href@noop {} {\bibfield  {journal} {\bibinfo  {journal} {arXiv:2111.13124}\
  } (\bibinfo {year} {2021})}\BibitemShut {NoStop}%
\bibitem [{\citenamefont {Cicconetti}\ \emph {et~al.}(2022)\citenamefont
  {Cicconetti}, \citenamefont {Conti},\ and\ \citenamefont
  {Passarella}}]{Cicco}%
  \BibitemOpen
  \bibfield  {author} {\bibinfo {author} {\bibfnamefont {C.}~\bibnamefont
  {Cicconetti}}, \bibinfo {author} {\bibfnamefont {M.}~\bibnamefont {Conti}},\
  and\ \bibinfo {author} {\bibfnamefont {A.}~\bibnamefont {Passarella}},\
  }\bibfield  {title} {\bibinfo {title} {Quality of service in quantum
  networks},\ }\href@noop {} {\bibfield  {journal} {\bibinfo  {journal} {IEEE
  Network}\ }\textbf {\bibinfo {volume} {36}},\ \bibinfo {pages} {24} (\bibinfo
  {year} {2022})}\BibitemShut {NoStop}%
\bibitem [{\citenamefont {Munro}\ \emph {et~al.}(2012)\citenamefont {Munro},
  \citenamefont {Stephens}, \citenamefont {Devitt}, \citenamefont {Harrison},\
  and\ \citenamefont {Nemoto}}]{QEC2}%
  \BibitemOpen
  \bibfield  {author} {\bibinfo {author} {\bibfnamefont {W.~J.}\ \bibnamefont
  {Munro}}, \bibinfo {author} {\bibfnamefont {A.~M.}\ \bibnamefont {Stephens}},
  \bibinfo {author} {\bibfnamefont {S.~J.}\ \bibnamefont {Devitt}}, \bibinfo
  {author} {\bibfnamefont {K.~A.}\ \bibnamefont {Harrison}},\ and\ \bibinfo
  {author} {\bibfnamefont {K.}~\bibnamefont {Nemoto}},\ }\bibfield  {title}
  {\bibinfo {title} {Quantum communication without the necessity of quantum
  memories},\ }\href@noop {} {\bibfield  {journal} {\bibinfo  {journal} {Nature
  Photonics}\ }\textbf {\bibinfo {volume} {6}},\ \bibinfo {pages} {777 }
  (\bibinfo {year} {2012})}\BibitemShut {NoStop}%
\bibitem [{\citenamefont {Abana}\ \emph {et~al.}(2024)\citenamefont {Abana},
  \citenamefont {Cubeddu}, \citenamefont {Man},\ and\ \citenamefont
  {Battou}}]{QROUTING1}%
  \BibitemOpen
  \bibfield  {author} {\bibinfo {author} {\bibfnamefont {A.}~\bibnamefont
  {Abana}}, \bibinfo {author} {\bibfnamefont {M.}~\bibnamefont {Cubeddu}},
  \bibinfo {author} {\bibfnamefont {V.~S.}\ \bibnamefont {Man}},\ and\ \bibinfo
  {author} {\bibfnamefont {A.}~\bibnamefont {Battou}},\ }\bibfield  {title}
  {\bibinfo {title} {Entanglement routing in quantum networks: A comprehensive
  survey},\ }\href@noop {} {\bibfield  {journal} {\bibinfo  {journal}
  {arXiv:2408.01234}\ } (\bibinfo {year} {2024})}\BibitemShut {NoStop}%
\bibitem [{\citenamefont {Li}\ \emph {et~al.}(2022)\citenamefont {Li},
  \citenamefont {Wang}, \citenamefont {jia}, \citenamefont {Sue}, \citenamefont
  {Yu}, \citenamefont {Sun},\ and\ \citenamefont {Lu}}]{QROUTING2}%
  \BibitemOpen
  \bibfield  {author} {\bibinfo {author} {\bibfnamefont {J.}~\bibnamefont
  {Li}}, \bibinfo {author} {\bibfnamefont {M.}~\bibnamefont {Wang}}, \bibinfo
  {author} {\bibfnamefont {Q.}~\bibnamefont {jia}}, \bibinfo {author}
  {\bibfnamefont {K.}~\bibnamefont {Sue}}, \bibinfo {author} {\bibfnamefont
  {N.}~\bibnamefont {Yu}}, \bibinfo {author} {\bibfnamefont {Q.}~\bibnamefont
  {Sun}},\ and\ \bibinfo {author} {\bibfnamefont {J.}~\bibnamefont {Lu}},\
  }\bibfield  {title} {\bibinfo {title} {Fidelity-guarantee entanglement
  routing in quantum networks},\ }\href@noop {} {\bibfield  {journal} {\bibinfo
   {journal} {arXiv:2111.07764v4}\ } (\bibinfo {year} {2022})}\BibitemShut
  {NoStop}%
\bibitem [{\citenamefont {Li}\ \emph {et~al.}(2023)\citenamefont {Li},
  \citenamefont {Chaudhary}, \citenamefont {Sanchez~Garcia},\ and\
  \citenamefont {Chowdhury}}]{QROUTING3}%
  \BibitemOpen
  \bibfield  {author} {\bibinfo {author} {\bibfnamefont {K.}~\bibnamefont
  {Li}}, \bibinfo {author} {\bibfnamefont {V.}~\bibnamefont {Chaudhary}},
  \bibinfo {author} {\bibfnamefont {S.}~\bibnamefont {Sanchez~Garcia}},\ and\
  \bibinfo {author} {\bibfnamefont {K.~R.}\ \bibnamefont {Chowdhury}},\
  }\bibfield  {title} {\bibinfo {title} {Q-firm: Fidelity-based rate maximizing
  routes for quantum networks},\ }\href@noop {} {\  (\bibinfo {year}
  {2023})}\BibitemShut {NoStop}%
\bibitem [{\citenamefont {Ralph}\ \emph {et~al.}(2005)\citenamefont {Ralph},
  \citenamefont {Hayes},\ and\ \citenamefont {Gilchrist}}]{redundancy_code}%
  \BibitemOpen
  \bibfield  {author} {\bibinfo {author} {\bibfnamefont {T.~C.}\ \bibnamefont
  {Ralph}}, \bibinfo {author} {\bibfnamefont {A.~J.~F.}\ \bibnamefont
  {Hayes}},\ and\ \bibinfo {author} {\bibfnamefont {A.}~\bibnamefont
  {Gilchrist}},\ }\bibfield  {title} {\bibinfo {title} {Loss-tolerant optical
  qubits},\ }\href@noop {} {\bibfield  {journal} {\bibinfo  {journal} {Phys.
  Rev. Lett.}\ }\textbf {\bibinfo {volume} {95}},\ \bibinfo {pages} {100501}
  (\bibinfo {year} {2005})}\BibitemShut {NoStop}%
\bibitem [{\citenamefont {Fowler}\ \emph
  {et~al.}(2010{\natexlab{a}})\citenamefont {Fowler}, \citenamefont {Wang},
  \citenamefont {Hill}, \citenamefont {Ladd}, \citenamefont {Van~Meter},\ and\
  \citenamefont {Hollenberg}}]{surfacecode}%
  \BibitemOpen
  \bibfield  {author} {\bibinfo {author} {\bibfnamefont {A.~G.}\ \bibnamefont
  {Fowler}}, \bibinfo {author} {\bibfnamefont {D.~S.}\ \bibnamefont {Wang}},
  \bibinfo {author} {\bibfnamefont {C.~D.}\ \bibnamefont {Hill}}, \bibinfo
  {author} {\bibfnamefont {T.~D.}\ \bibnamefont {Ladd}}, \bibinfo {author}
  {\bibfnamefont {R.}~\bibnamefont {Van~Meter}},\ and\ \bibinfo {author}
  {\bibfnamefont {L.~C.~L.}\ \bibnamefont {Hollenberg}},\ }\bibfield  {title}
  {\bibinfo {title} {Surface code quantum communication},\ }\href@noop {}
  {\bibfield  {journal} {\bibinfo  {journal} {Phys. Rev. Lett.}\ }\textbf
  {\bibinfo {volume} {104}},\ \bibinfo {pages} {180503} (\bibinfo {year}
  {2010}{\natexlab{a}})}\BibitemShut {NoStop}%
\bibitem [{\citenamefont {Jiang}\ \emph {et~al.}(2009)\citenamefont {Jiang},
  \citenamefont {Taylor}, \citenamefont {Nemoto}, \citenamefont {Munro},
  \citenamefont {Van~Meter},\ and\ \citenamefont {Lukin}}]{QEC1}%
  \BibitemOpen
  \bibfield  {author} {\bibinfo {author} {\bibfnamefont {L.}~\bibnamefont
  {Jiang}}, \bibinfo {author} {\bibfnamefont {J.~M.}\ \bibnamefont {Taylor}},
  \bibinfo {author} {\bibfnamefont {K.}~\bibnamefont {Nemoto}}, \bibinfo
  {author} {\bibfnamefont {W.~J.}\ \bibnamefont {Munro}}, \bibinfo {author}
  {\bibfnamefont {R.}~\bibnamefont {Van~Meter}},\ and\ \bibinfo {author}
  {\bibfnamefont {M.~D.}\ \bibnamefont {Lukin}},\ }\bibfield  {title} {\bibinfo
  {title} {Quantum repeater with encoding},\ }\href@noop {} {\bibfield
  {journal} {\bibinfo  {journal} {Phys. Rev. A}\ ,\ \bibinfo {pages} {032325}}
  (\bibinfo {year} {2009})}\BibitemShut {NoStop}%
\bibitem [{\citenamefont {Gottesman}\ \emph {et~al.}(2001)\citenamefont
  {Gottesman}, \citenamefont {Kitaev},\ and\ \citenamefont {Preskill}}]{GKP1}%
  \BibitemOpen
  \bibfield  {author} {\bibinfo {author} {\bibfnamefont {D.}~\bibnamefont
  {Gottesman}}, \bibinfo {author} {\bibfnamefont {A.}~\bibnamefont {Kitaev}},\
  and\ \bibinfo {author} {\bibfnamefont {J.}~\bibnamefont {Preskill}},\
  }\bibfield  {title} {\bibinfo {title} {Encoding a qubit in an oscillator},\
  }\href@noop {} {\bibfield  {journal} {\bibinfo  {journal} {Phys. Rev. A}\
  }\textbf {\bibinfo {volume} {64}},\ \bibinfo {pages} {012310} (\bibinfo
  {year} {2001})}\BibitemShut {NoStop}%
\bibitem [{\citenamefont {Fowler}\ \emph
  {et~al.}(2010{\natexlab{b}})\citenamefont {Fowler}, \citenamefont {Wang},
  \citenamefont {Hill}, \citenamefont {Ladd}, \citenamefont {Van~Meter},\ and\
  \citenamefont {Hollenberg}}]{QEC3}%
  \BibitemOpen
  \bibfield  {author} {\bibinfo {author} {\bibfnamefont {A.~G.}\ \bibnamefont
  {Fowler}}, \bibinfo {author} {\bibfnamefont {D.~S.}\ \bibnamefont {Wang}},
  \bibinfo {author} {\bibfnamefont {C.~H.}\ \bibnamefont {Hill}}, \bibinfo
  {author} {\bibfnamefont {T.~D.}\ \bibnamefont {Ladd}}, \bibinfo {author}
  {\bibfnamefont {R.}~\bibnamefont {Van~Meter}},\ and\ \bibinfo {author}
  {\bibfnamefont {C.~L.}\ \bibnamefont {Hollenberg}},\ }\bibfield  {title}
  {\bibinfo {title} {Surface code quantum communication},\ }\href@noop {}
  {\bibfield  {journal} {\bibinfo  {journal} {Phys. Rev. Lett.}\ }\textbf
  {\bibinfo {volume} {104}},\ \bibinfo {pages} {180503} (\bibinfo {year}
  {2010}{\natexlab{b}})}\BibitemShut {NoStop}%
\bibitem [{\citenamefont {Azuma}\ \emph {et~al.}(2015)\citenamefont {Azuma},
  \citenamefont {Tamaki},\ and\ \citenamefont {Lo}}]{QEC4}%
  \BibitemOpen
  \bibfield  {author} {\bibinfo {author} {\bibfnamefont {K.}~\bibnamefont
  {Azuma}}, \bibinfo {author} {\bibfnamefont {K.}~\bibnamefont {Tamaki}},\ and\
  \bibinfo {author} {\bibfnamefont {H.~K.}\ \bibnamefont {Lo}},\ }\bibfield
  {title} {\bibinfo {title} {All-photonic quantum repeaters},\ }\href@noop {}
  {\bibfield  {journal} {\bibinfo  {journal} {Nat. Commun.}\ }\textbf {\bibinfo
  {volume} {6}},\ \bibinfo {pages} {6787} (\bibinfo {year} {2015})}\BibitemShut
  {NoStop}%
\bibitem [{\citenamefont {Mulidharan}\ \emph {et~al.}(2014)\citenamefont
  {Mulidharan}, \citenamefont {Kim}, \citenamefont {Lutkenhaus}, \citenamefont
  {Lucian},\ and\ \citenamefont {Jiang}}]{QEC5}%
  \BibitemOpen
  \bibfield  {author} {\bibinfo {author} {\bibfnamefont {S.}~\bibnamefont
  {Mulidharan}}, \bibinfo {author} {\bibfnamefont {J.}~\bibnamefont {Kim}},
  \bibinfo {author} {\bibfnamefont {N.}~\bibnamefont {Lutkenhaus}}, \bibinfo
  {author} {\bibfnamefont {M.~D.}\ \bibnamefont {Lucian}},\ and\ \bibinfo
  {author} {\bibfnamefont {L.}~\bibnamefont {Jiang}},\ }\bibfield  {title}
  {\bibinfo {title} {Ultrafast and fault-tolerant quantum communication across
  long distances},\ }\href@noop {} {\bibfield  {journal} {\bibinfo  {journal}
  {Phys. Rev. Lett.}\ }\textbf {\bibinfo {volume} {112}},\ \bibinfo {pages}
  {250501} (\bibinfo {year} {2014})}\BibitemShut {NoStop}%
\bibitem [{\citenamefont {Lo~Piparo}\ \emph {et~al.}(2024)\citenamefont
  {Lo~Piparo}, \citenamefont {Munro},\ and\ \citenamefont
  {Nemoto}}]{temporal_delay}%
  \BibitemOpen
  \bibfield  {author} {\bibinfo {author} {\bibfnamefont {N.}~\bibnamefont
  {Lo~Piparo}}, \bibinfo {author} {\bibfnamefont {W.~J.}\ \bibnamefont
  {Munro}},\ and\ \bibinfo {author} {\bibfnamefont {K.}~\bibnamefont
  {Nemoto}},\ }\bibfield  {title} {\bibinfo {title} {Quantum aggregation with
  temporal delay},\ }\href@noop {} {\bibfield  {journal} {\bibinfo  {journal}
  {Phys. Rev. A}\ }\textbf {\bibinfo {volume} {110}},\ \bibinfo {pages}
  {032613} (\bibinfo {year} {2024})}\BibitemShut {NoStop}%
\bibitem [{ref()}]{reference_note1}%
  \BibitemOpen
  \href@noop {} {\emph {\bibinfo {title} {\textnormal{Based on these
  observations, we generalize an aggregated path as a set of paths connecting
  two nodes with varying delays. Accordingly, we refine the definition of delay
  for such paths. For intuition, assume temporal delays are negligible, so
  delay depends only on path length. According to the protocol, if enough
  resources arrive via the shortest path, the receiver can decode the state
  immediatelyÑyielding a delay equal to that of the shortest path. However, if
  additional resources are needed, decoding must wait for longer paths. Thus,
  the delay of an aggregated path is defined as the weighted average of the
  delays of all contributing paths, with weights given by the probability that
  each path successfully delivers sufficient resources.}}}}\BibitemShut {Stop}%
\bibitem [{\citenamefont {Bunandar}\ \emph
  {et~al.}(2018{\natexlab{b}})\citenamefont {Bunandar}, \citenamefont
  {Lentine}, \citenamefont {Lee}, \citenamefont {Cai}, \citenamefont {Long},
  \citenamefont {Boynton}, \citenamefont {Martinez}, \citenamefont {DeRose},
  \citenamefont {Chen}, \citenamefont {Grein}, \citenamefont {Trotter},
  \citenamefont {Starbuck}, \citenamefont {Pomerene}, \citenamefont {Hamilton},
  \citenamefont {Wong}, \citenamefont {Camacho}, \citenamefont {Davids},
  \citenamefont {Urayama},\ and\ \citenamefont {Englund}}]{QKD_adv}%
  \BibitemOpen
  \bibfield  {author} {\bibinfo {author} {\bibfnamefont {D.}~\bibnamefont
  {Bunandar}}, \bibinfo {author} {\bibfnamefont {A.}~\bibnamefont {Lentine}},
  \bibinfo {author} {\bibfnamefont {C.}~\bibnamefont {Lee}}, \bibinfo {author}
  {\bibfnamefont {H.}~\bibnamefont {Cai}}, \bibinfo {author} {\bibfnamefont
  {C.~M.}\ \bibnamefont {Long}}, \bibinfo {author} {\bibfnamefont
  {N.}~\bibnamefont {Boynton}}, \bibinfo {author} {\bibfnamefont
  {N.}~\bibnamefont {Martinez}}, \bibinfo {author} {\bibfnamefont
  {C.}~\bibnamefont {DeRose}}, \bibinfo {author} {\bibfnamefont
  {C.}~\bibnamefont {Chen}}, \bibinfo {author} {\bibfnamefont {M.}~\bibnamefont
  {Grein}}, \bibinfo {author} {\bibfnamefont {A.}~\bibnamefont {Trotter}},
  \bibinfo {author} {\bibfnamefont {A.}~\bibnamefont {Starbuck}}, \bibinfo
  {author} {\bibfnamefont {A.}~\bibnamefont {Pomerene}}, \bibinfo {author}
  {\bibfnamefont {S.}~\bibnamefont {Hamilton}}, \bibinfo {author}
  {\bibfnamefont {F.~N.~C.}\ \bibnamefont {Wong}}, \bibinfo {author}
  {\bibfnamefont {R.}~\bibnamefont {Camacho}}, \bibinfo {author} {\bibfnamefont
  {P.}~\bibnamefont {Davids}}, \bibinfo {author} {\bibfnamefont
  {J.}~\bibnamefont {Urayama}},\ and\ \bibinfo {author} {\bibfnamefont
  {D.}~\bibnamefont {Englund}},\ }\bibfield  {title} {\bibinfo {title}
  {Metropolitan quantum key distribution with silicon photonics},\ }\href@noop
  {} {\bibfield  {journal} {\bibinfo  {journal} {Phys. Rev. X}\ }\textbf
  {\bibinfo {volume} {8}},\ \bibinfo {pages} {021009} (\bibinfo {year}
  {2018}{\natexlab{b}})}\BibitemShut {NoStop}%
\bibitem [{\citenamefont {Devitt}\ \emph {et~al.}(2013)\citenamefont {Devitt},
  \citenamefont {Stephens}, \citenamefont {Munro},\ and\ \citenamefont
  {Nemoto}}]{QComp_Bill}%
  \BibitemOpen
  \bibfield  {author} {\bibinfo {author} {\bibfnamefont {S.~J.}\ \bibnamefont
  {Devitt}}, \bibinfo {author} {\bibfnamefont {A.~M.}\ \bibnamefont
  {Stephens}}, \bibinfo {author} {\bibfnamefont {W.~J.}\ \bibnamefont
  {Munro}},\ and\ \bibinfo {author} {\bibfnamefont {K.}~\bibnamefont
  {Nemoto}},\ }\bibfield  {title} {\bibinfo {title} {Requirements for
  fault-tolerant factoring on an atom-optics quantum computer},\ }\href@noop {}
  {\bibfield  {journal} {\bibinfo  {journal} {Nature Communications}\ }\textbf
  {\bibinfo {volume} {4}},\ \bibinfo {pages} {2524} (\bibinfo {year}
  {2013})}\BibitemShut {NoStop}%
\bibitem [{\citenamefont {Panayi}\ \emph {et~al.}(2014)\citenamefont {Panayi},
  \citenamefont {Razavi}, \citenamefont {Ma},\ and\ \citenamefont
  {Lutkenhaus}}]{memory_ass}%
  \BibitemOpen
  \bibfield  {author} {\bibinfo {author} {\bibfnamefont {C.}~\bibnamefont
  {Panayi}}, \bibinfo {author} {\bibfnamefont {M.}~\bibnamefont {Razavi}},
  \bibinfo {author} {\bibfnamefont {X.}~\bibnamefont {Ma}},\ and\ \bibinfo
  {author} {\bibfnamefont {N.}~\bibnamefont {Lutkenhaus}},\ }\bibfield  {title}
  {\bibinfo {title} {Memory-assisted measurement-device-independent quantum key
  distribution},\ }\href@noop {} {\bibfield  {journal} {\bibinfo  {journal}
  {New J. Phys.}\ }\textbf {\bibinfo {volume} {16}},\ \bibinfo {pages} {043005}
  (\bibinfo {year} {2014})}\BibitemShut {NoStop}%
\bibitem [{\citenamefont {Zhou}\ \emph
  {et~al.}(2018{\natexlab{a}})\citenamefont {Zhou}, \citenamefont {Zhang},
  \citenamefont {Preskill},\ and\ \citenamefont {Jiang}}]{QImaging4}%
  \BibitemOpen
  \bibfield  {author} {\bibinfo {author} {\bibfnamefont {S.}~\bibnamefont
  {Zhou}}, \bibinfo {author} {\bibfnamefont {M.}~\bibnamefont {Zhang}},
  \bibinfo {author} {\bibfnamefont {J.}~\bibnamefont {Preskill}},\ and\
  \bibinfo {author} {\bibfnamefont {L.}~\bibnamefont {Jiang}},\ }\bibfield
  {title} {\bibinfo {title} {Achieving the heisenberg limit in quantum
  metrology using quantum error correction},\ }\href@noop {} {\bibfield
  {journal} {\bibinfo  {journal} {Nat. Commun.}\ }\textbf {\bibinfo {volume}
  {9}},\ \bibinfo {pages} {78} (\bibinfo {year}
  {2018}{\natexlab{a}})}\BibitemShut {NoStop}%
\bibitem [{\citenamefont {Zhou}\ \emph
  {et~al.}(2018{\natexlab{b}})\citenamefont {Zhou}, \citenamefont {Zhang},
  \citenamefont {Preskill},\ and\ \citenamefont {Jiang}}]{Heisenberg_Lim}%
  \BibitemOpen
  \bibfield  {author} {\bibinfo {author} {\bibfnamefont {S.}~\bibnamefont
  {Zhou}}, \bibinfo {author} {\bibfnamefont {M.}~\bibnamefont {Zhang}},
  \bibinfo {author} {\bibfnamefont {J.}~\bibnamefont {Preskill}},\ and\
  \bibinfo {author} {\bibfnamefont {L.}~\bibnamefont {Jiang}},\ }\bibfield
  {title} {\bibinfo {title} {Achieving the heisenberg limit in quantum
  metrology using quantum error correction},\ }\href@noop {} {\bibfield
  {journal} {\bibinfo  {journal} {Nature Communications}\ }\textbf {\bibinfo
  {volume} {9}},\ \bibinfo {pages} {78} (\bibinfo {year}
  {2018}{\natexlab{b}})}\BibitemShut {NoStop}%
\bibitem [{\citenamefont {Elliott}(2002)}]{sync1}%
  \BibitemOpen
  \bibfield  {author} {\bibinfo {author} {\bibfnamefont {C.}~\bibnamefont
  {Elliott}},\ }\bibfield  {title} {\bibinfo {title} {Building the quantum
  network},\ }\href@noop {} {\bibfield  {journal} {\bibinfo  {journal} {New J.
  Phys.}\ }\textbf {\bibinfo {volume} {4}},\ \bibinfo {pages} {46} (\bibinfo
  {year} {2002})}\BibitemShut {NoStop}%
\bibitem [{\citenamefont {Korzh}\ \emph {et~al.}(2015)\citenamefont {Korzh},
  \citenamefont {Lim}, \citenamefont {Houlmann}, \citenamefont {Gisin},
  \citenamefont {J.}, \citenamefont {Nolan}, \citenamefont {B.}, \citenamefont
  {Thew},\ and\ \citenamefont {H.}}]{sync2}%
  \BibitemOpen
  \bibfield  {author} {\bibinfo {author} {\bibfnamefont {B.}~\bibnamefont
  {Korzh}}, \bibinfo {author} {\bibfnamefont {C.~W.}\ \bibnamefont {Lim}},
  \bibinfo {author} {\bibfnamefont {R.}~\bibnamefont {Houlmann}}, \bibinfo
  {author} {\bibfnamefont {N.}~\bibnamefont {Gisin}}, \bibinfo {author}
  {\bibfnamefont {L.~M.}\ \bibnamefont {J.}}, \bibinfo {author} {\bibfnamefont
  {D.}~\bibnamefont {Nolan}}, \bibinfo {author} {\bibfnamefont
  {S.}~\bibnamefont {B.}}, \bibinfo {author} {\bibfnamefont {R.}~\bibnamefont
  {Thew}},\ and\ \bibinfo {author} {\bibfnamefont {Z.}~\bibnamefont {H.}},\
  }\bibfield  {title} {\bibinfo {title} {Provably secure and practical quantum
  key distribution over 307 km of optical fiber},\ }\href@noop {} {\bibfield
  {journal} {\bibinfo  {journal} {Nature Photonics}\ }\textbf {\bibinfo
  {volume} {9}},\ \bibinfo {pages} {163} (\bibinfo {year} {2015})}\BibitemShut
  {NoStop}%
\bibitem [{\citenamefont {Van~Meter}\ and\ \citenamefont
  {Horsman}(2013)}]{Qcomp_Rod}%
  \BibitemOpen
  \bibfield  {author} {\bibinfo {author} {\bibfnamefont {R.}~\bibnamefont
  {Van~Meter}}\ and\ \bibinfo {author} {\bibfnamefont {D.}~\bibnamefont
  {Horsman}},\ }\bibfield  {title} {\bibinfo {title} {A blueprint for building
  a quantum computer},\ }\href@noop {} {\bibfield  {journal} {\bibinfo
  {journal} {Communications of the ACM}\ }\textbf {\bibinfo {volume} {10}},\
  \bibinfo {pages} {83} (\bibinfo {year} {2013})}\BibitemShut {NoStop}%
\bibitem [{\citenamefont {Grassl}\ \emph {et~al.}(1999)\citenamefont {Grassl},
  \citenamefont {Geiselmann},\ and\ \citenamefont {Beth}}]{Reed_Sal}%
  \BibitemOpen
  \bibfield  {author} {\bibinfo {author} {\bibfnamefont {M.}~\bibnamefont
  {Grassl}}, \bibinfo {author} {\bibfnamefont {W.}~\bibnamefont {Geiselmann}},\
  and\ \bibinfo {author} {\bibfnamefont {T.}~\bibnamefont {Beth}},\ }\bibfield
  {title} {\bibinfo {title} {Loss-tolerant optical qubits},\ }\href@noop {}
  {\bibfield  {journal} {\bibinfo  {journal} {International Symposium on
  Applied Algebra, Algebraic Algorithms, and Error-Correcting Codes,}\ ,\
  \bibinfo {pages} {231}} (\bibinfo {year} {1999})}\BibitemShut {NoStop}%
\bibitem [{\citenamefont {Akyildiz}\ \emph {et~al.}(2005)\citenamefont
  {Akyildiz}, \citenamefont {Wang},\ and\ \citenamefont {Wang}}]{MeshN1}%
  \BibitemOpen
  \bibfield  {author} {\bibinfo {author} {\bibfnamefont {I.~F.}\ \bibnamefont
  {Akyildiz}}, \bibinfo {author} {\bibfnamefont {X.}~\bibnamefont {Wang}},\
  and\ \bibinfo {author} {\bibfnamefont {W.}~\bibnamefont {Wang}},\ }\bibfield
  {title} {\bibinfo {title} {Wireless mesh networks: A survey},\ }\href@noop {}
  {\bibfield  {journal} {\bibinfo  {journal} {Computer Networks}\ }\textbf
  {\bibinfo {volume} {47}},\ \bibinfo {pages} {445} (\bibinfo {year}
  {2005})}\BibitemShut {NoStop}%
\bibitem [{\citenamefont {Liu}\ and\ \citenamefont {El~Zarki}(2005)}]{MeshN2}%
  \BibitemOpen
  \bibfield  {author} {\bibinfo {author} {\bibfnamefont {K.}~\bibnamefont
  {Liu}}\ and\ \bibinfo {author} {\bibfnamefont {M.}~\bibnamefont {El~Zarki}},\
  }\bibfield  {title} {\bibinfo {title} {Qos routing in wireless ad hoc
  networks},\ }\href@noop {} {\bibfield  {journal} {\bibinfo  {journal} {IEEE
  Network}\ }\textbf {\bibinfo {volume} {19(3)}},\ \bibinfo {pages} {12}
  (\bibinfo {year} {2005})}\BibitemShut {NoStop}%
\end{thebibliography}%

\appendix

\section{Fidelities used in the unencoded case}

Here we list the analytical expressions for the fidelities used to
determine the curves in Fig. \ref{fig:Fidelities-of-the-unencoded}.
We denote with $F_{i+j}$ the fidelity of the corresponding $i+j$
configuration, in which $i$ qudits are traveling in path 1 and $j$
qudits are traveling in path 2.

$F_{5+0}=p_{1}^{5}.$

$\begin{array}{cc}
F_{4+1}= & \negmedspace\negmedspace\negmedspace\negmedspace\negmedspace\negmedspace\negmedspace\negmedspace\negmedspace\negmedspace\negmedspace\negmedspace\negmedspace\negmedspace p_{1}^{4}p_{2}\left[\left(1-p_{d}\right)^{4}+\frac{4}{7}p_{d}\left(1-p_{d}\right)^{3}\right.\\
 & \left.+\frac{6}{7^{2}}p_{d}^{2}\left(1-p_{d}\right)^{2}+\frac{4}{7^{3}}p_{d}^{3}\left(1-p_{d}\right)+p^{4}/7^{4}\right].
\end{array}$

$\begin{array}{cc}
F_{3+2}= & p_{1}^{3}p_{2}^{2}\left(1-p_{d}\right)^{3}+\frac{3}{7}p_{d}\left(1-p_{d}\right)^{2}\\
 & \negmedspace\negmedspace\negmedspace\negmedspace\negmedspace\negmedspace\negmedspace\negmedspace\negmedspace\negmedspace\negmedspace\negmedspace\negmedspace\left.+\frac{3}{7^{2}}p_{d}^{2}\left(1-p_{d}\right)+p^{3}/7^{3}\right].
\end{array}$

$F_{2+3}=p_{1}^{2}p_{2}^{3}\left[\left(1-p_{d}\right)^{2}+\frac{2}{7}p_{d}\left(1-p_{d}\right)+p^{2}/7^{2}\right].$

$F_{1+4}=p_{1}p_{2}^{4}\left[\left(1-p_{d}\right)+p/7\right]$.

$F_{0+5}=p_{2}^{5}.$
\end{document}